\documentclass[12pt]{article}

\usepackage{booktabs}
\usepackage{color}
\renewcommand{\baselinestretch}{1.5}
\usepackage{amsthm,amsmath, amssymb, amsfonts, amscd, xspace, pifont, color, bm}
\usepackage{mathtools,etoolbox}
\usepackage{epsfig, psfrag}
\usepackage{float, fancybox, fullpage}
\usepackage{delarray}
\usepackage{hyperref}
\usepackage{url}
\usepackage{anysize}
\usepackage{multirow}
\usepackage{geometry}
\usepackage{rotating}
\usepackage{graphicx}
\usepackage{caption}
\usepackage{subcaption}
\usepackage[round]{natbib} 
\usepackage[T1]{fontenc}
\usepackage[utf8]{inputenc}
\usepackage{authblk}
\usepackage{xcolor}
\usepackage{algorithm}
\usepackage{algpseudocode}
\usepackage{pifont}
\usepackage[algo2e]{algorithm2e} 
\usepackage{enumitem}
\usepackage[short]{optidef}
\usepackage{physics}
\usepackage{relsize}
\usepackage[export]{adjustbox}
\usepackage{caption}
\usepackage{subcaption}
\usepackage{xparse}
\usepackage{dsfont}
\usepackage{siunitx}
\usepackage{nicematrix,tikz}
\usepackage{arydshln}
\usepackage{tabularray}
\usepackage{array}
\usepackage{ragged2e}
\usepackage{tabularx}
\usepackage{booktabs}
\usepackage{makecell}

\newcommand{\ze}{Z\kern-0.45emZ}
\newcommand{\esp}{I\kern-0.37emE}
\newcommand{\N}{I\kern-0.37emN}
\newcommand{\one}{ {\rm 1\kern-0.19eml} }
\newcommand{\realset}{I\kern-0.37emR}

\newcommand{\bx}{\boldsymbol{x}}

\newcommand{\bX}{\boldsymbol{X}}

\newcommand{\bw}{\boldsymbol{w}}
\newcommand{\bv}{\boldsymbol{v}}
\newcommand{\bu}{\boldsymbol{u}}
\newcommand{\bs}{\boldsymbol{s}}
\newcommand{\bt}{\boldsymbol{t}}
\newcommand{\be}{\boldsymbol{e}}

\newcommand{\bb}{\boldsymbol{b}}
\newcommand{\bz}{\boldsymbol{z}}

\newcommand{\bd}{\boldsymbol{d}}
\newcommand{\bc}{\boldsymbol{c}}
\newcommand{\floor}[1]{\left\lfloor #1 \right\rfloor}

\newcommand{\bbeta}{\mbox{\boldmath $\beta$}}
\newcommand{\bbetazero}{\mbox{$\beta_0$}}

\newcommand{\bSigma}{\mbox{\boldmath $\Sigma$}}

\DeclarePairedDelimiter{\normtwo}{\lVert}{\rVert} 

\SetCommentSty{mycommfont}

\makeatletter
\newcommand{\ssymbol}[1]{^{\@fnsymbol{#1}}}
\makeatother

\NewDocumentCommand{\INTERVALINNARDS}{ m m }{
    #1 {,} #2
}
\NewDocumentCommand{\interval}{ s m >{\SplitArgument{1}{,}}m m o }{
    \IfBooleanTF{#1}{
        \left#2 \INTERVALINNARDS #3 \right#4
    }{
        \IfValueTF{#5}{
            #5{#2} \INTERVALINNARDS #3 #5{#4}
        }{
            #2 \INTERVALINNARDS #3 #4
        }
    }
}

\SetKwInput{KwInput}{Input}                
\SetKwInput{KwOutput}{Output}              

\newtheorem{theorem}{Theorem}

\def\argmax{\mathop{\rm argmax}}

\DeclarePairedDelimiterX\Card[1]\lvert\rvert{
  \ifblank{#1}{{-}}{#1}
}
\DeclarePairedDelimiter{\norm1}{\lVert}{\rVert} 

\renewcommand{\vec}[1]{\bm{\mathrm{#1}}}

\def\red{\textcolor{black}}

\newcommand{\blind}{0}

\begin{document}

\def\spacingset#1{\renewcommand{\baselinestretch}%
{#1}\small\normalsize} \spacingset{1}


\title{Sparse Learning and Class Probability Estimation with Weighted Support Vector Machines}
\date{}

\author[1]{Liyun Zeng}
\author[1,2,*]{Hao Helen Zhang}
\affil[1]{Statistics and Data Science GIDP, University of Arizona, Tucson, AZ, 85721, USA}
\affil[2]{Department of Mathematics, University of Arizona, Tucson, AZ, 85721, USA}
\affil[*]{Corresponding author: haozhang@arizona.edu}

\maketitle

\if1\blind
{
  \bigskip
  \bigskip
  \bigskip
  \begin{center}
    {\LARGE\bf Title}
\end{center}
  \medskip
} \fi

\bigskip
\begin{abstract}
Classification and probability estimation are fundamental tasks with broad applications across modern machine learning and data science, spanning fields such as biology, medicine, engineering, and computer science. The recent development of weighted Support Vector Machines (wSVMs) has demonstrated considerable promise in robustly and accurately predicting class probabilities and performing classification across a variety of problems (Wang et al., 2008). However, the existing framework relies on an $\ell^2$-norm regularized binary wSVMs optimization formulation, which is designed for dense features and exhibits limited performance in the presence of sparse features with redundant noise, a common challenge in real applications. Effective sparse learning thus requires prescreening of important variables for each binary wSVM to ensure accurate estimation of pairwise conditional probabilities. In this paper, we propose a novel class of wSVMs frameworks that incorporate automatic variable selection with accurate probability estimation for sparse learning problems. We developed efficient algorithms for variable selection by solving either the $\ell^1$-norm or elastic net regularized binary wSVMs optimization problems. Class probability is then estimated either via the $\ell^2$-norm regularized wSVMs framework applied to the selected variables, or directly through elastic net regularized wSVMs. The two-step approach offers a strong advantage in simultaneous automatic variable selection and reliable probability estimators with competitive computational efficiency. The elastic net regularized wSVMs achieve superior performance in both variable selection and probability estimation, with the added benefit of variable grouping, at the cost of increases compensation time for high dimensional settings. The proposed wSVMs-based sparse learning methods are broadly applicable and can be naturally extended to $K$-class problems through ensemble learning.
\end{abstract}

\noindent%
\vfill

\newpage
\spacingset{1.75} 

\section{Introduction}
Binary classification is a foundational topic in machine learning and artificial intelligence  \citep{Bishop1162264}. It serves as the building block for multiclass classification, where decision boundaries are constructed by aggregating multiple binary classification rules through approaches such as pairwise coupling and One-vs-All \citep{Hastie2009}. It underpins a wide range of real-world applications, including disease diagnosis to determine whether a patient has a particular condition, industrial quality control to assess whether a product meets a given specification, and information retrieval to decide whether a document belongs in a set of search results. In modern machine learning, it is common to encounter datasets with a large number of candidate predictive variables, and only a small subset of which carry meaningful predictive signal \citep{Li9088292}. In biomedical research, for instance, high-throughput technologies generate vast amounts of  genetic data whose dimensionality far exceeds the available sample size, yet the underlying disease mechanism is typically sparse and controlled by only a few key predictors \citep{Pedrobbk007, ye2012}.  \red{In such settings, classification methods that do not incorporate variable selection are prone to poor performance, as irrelevant and noisy features introduce overfitting and over-parametrization. The resulting models are not only too complex but also difficult to interpret due to the unnecessary inclusion of all features} \citep{Wang6126288, Rohini7887916}. It is therefore essential to identify the truly informative variables to simultaneously improve classification accuracy and enhance model interpretability.

Support vector machines (SVMs) are supervised learning models that perform accurate classification in high-dimensional data through geometric margin maximization. Their computational efficiency comes from dual optimization that depends on the sample size $n$ rather than data dimensionality $p$, making them particularly effective for high-dimensional problems \citep{cristianini_shawe-taylor_2000}. Standard soft-margin SVMs are equivalent to an optimization problem with the hinge loss and $\ell^2$-norm regularization \citep{lin_support_2002}. Given $n$ training samples $(\bx_1, y_i), (\bx_2, y_2), \ldots, (\bx_n, y_n)$, where $\bx_i \in \mathbb{R}^p$ represents the $p$-dimensional predictor vector and and $y \in \{-1, +1\}$ denotes the class label, SVMs learn the decision function $f(\bx) = \bbetazero + \bbeta^{\small\mathrm{T}}\bx$ by solving the following regularization problem:
\begin{equation}
        \min_{\bbetazero,\bbeta}\enskip\frac{1}{n}\sum_{i=1}^nL(y_i(\beta_0 + \bbeta^{\small\mathrm{T}}\bx_i))+\frac{\lambda}{2} \normtwo{\bbeta}_{2}^2 , \label{l2svm}
\end{equation}
where $\bbetazero \in \realset$ and $\bbeta \in \realset^p$, $\phi (\bx_i) = y_if(\bx_i)$ is the functional margin,  $L(z)=(1-z)_+=\max\{0, 1-z\}$ is the hinge loss, $\lambda > 0$ is the regularized term that controls model complexity and balances the bias–variance trade-off \citep{Hastie2009}.  \red{While the $\ell^2$-penalty shrinks coefficients towards zero to control} variance \red{in the presence} of highly correlated predictors, \red{standard SVMs retain all variables in the final classifier and do not perform} variable selection \citep{hastie_entire_2004, Hastie2009}. To address this limitation in sparse learning problems, \red{several sparse SVM variants have been proposed, including} the $\ell^1$-norm SVM \citep{Zhu04}, \red{SCAD-SVM \citep{zhang_gene_2006}}, and \red{ElasticNet SVM (DrSVM) \citep{wang_doubly_2006}}.

In particular, the linear $L_1$-SVM imposes the $\ell^1$-norm penalty in \ref{l2svm} and solves:
\begin{equation}
        \min_{\bbetazero,\bbeta}\enskip\frac{1}{n}\sum_{i=1}^nL(y_i(\beta_0 + \bbeta^{\small\mathrm{T}}\bx_i))+\lambda \normtwo{\bbeta}_{1} , \label{l1svm}
\end{equation}
The $\ell^1$-norm penalty performs Lasso-\red{type} variable selection \citep{Tibshirani02080} by shrinking small coefficients to exactly zero.  It outperforms standard $L_2$-SVM in high dimensional settings with noisy variables \citep{Donoho95waveletshrinkage, ng_feature_2004} and effectively improves prediction accuracy, it has two main limitations. \red{First}, when multiple correlated predictors are relevant to the response, $L_1$ regularization penalty arbitrarily selects a subset and discard \red{the rest. In gene pathway analysis \citep{reimand_pathway_2019}, for example, functionally related genes within a} pathway are highly correlated, yet $L_1$ penalty \red{may select only one gene}, whereas including the entire group would better support downstream analysis. Second, when $p \gg n$, the $L_1$ penalty can select at most $n$ predictors \citep{Rosset2004BoostingAA}. In clinical trials with sample sizes on the order of tens but thousands of gene biomarkers, this constraint limits identification of complex disease pathways. \cite{wang_doubly_2006} addressed these limitations by proposing the elastic net, combining $L_1$ and $L_2$ regularization in the doubly regularized (ElasticNet) SVM:
\begin{equation}
        \min_{\bbetazero,\bbeta}\enskip\frac{1}{n}\sum_{i=1}^nL(y_i(\beta_0 + \bbeta^{\small\mathrm{T}}\bx_i)) + \lambda_1 \normtwo{\bbeta}_{1} + \frac{\lambda_2}{2} \normtwo{\bbeta}_{2}^2, \label{elasticsvm}
\end{equation}
where $\lambda_1$ and $\lambda_2$ are tuning parameters that control model complexity. Like the $L_1$-SVM, the ElasticNet SVM performs automatic variable selection and continuous shrinkage. Additionally, it exhibits a ``grouping effect", selecting groups of correlated variables together by encouraging similar coefficient values. Unlike the $L_1$-SVM, the ElasticNet SVM can select more than $n$ variables. 

Standard SVMs classify new data $\bx$ using $\mbox{sign}(\bbetazero + \bbeta^{\small\mathrm{T}}\bx)$ and directly target the decision boundary. Consequently, they cannot predict class probabilities or provide prediction confidence measures, unlike soft classifiers such as naïve Bayes (NB), logistic regression, and linear discriminant analysis \citep{scholkopf_learning_2002, Hastie2009}. \citep{WSL2008} proposed weighted SVMs (wSVMs) for probability estimation, solving a sequence of wSVM classifiers to bracket class probabilities. However, $L_2$-wSVMs may perform poorly with redundant or noisy features, making variable selection essential for high dimensional data. We propose incorporating automatic variable selection into wSVMs to achieve sparse learning for class probability estimation and classification while preserving their probability prediction capabilities.

In this paper, motivated by $L_1$-SVMs for variable selection and ElaticNet SVMs for combined selection and grouping, we \red{introduce} sparse learning for weighted SVMs (wSVMs) that perform automatic variable selection and class probability estimation for high-dimensional settings with redundant and noisy features. Compared to the $L_2$-wSVMs, the proposed methods show improved variable selection, probability estimation, and classification performance.  These methods can be solved via linear programming (LP) and quadratic programming (QP) using R, Python, or MATLAB. Due to their divide-and-conquer nature, the algorithms have potential for parallel acceleration using CUDA, multi-core parallel computing, high-performance computing (HPC) clusters, or massively parallel computing (MPP) systems. 

The paper is organized as follows. Section \ref{sec:l2wsvmmethod} reviews the binary wSVM framework. Sections \ref{l1l2wsvm} and \ref{sec:elasticnetwsvm} present the new sparse learning frameworks and prove the grouping effect. Sections \ref{sec:complex} and \ref{sec:tuning} discuss computational complexity and hyperparameter tuning. Sections \ref{sec:simu} and 
\ref{sec:realdata} present simulations and real data analysis, followed by conclusions in Section \ref{sec:cond}.

\section{Binary Weighted SVMs}\label{sec:l2wsvmmethod}

\subsection{Notations} \label{sec2:notations}
Let the training set be $\left\{\left(\bx_i,y_i\right), i=1,\ldots,n\right\}$, where $\bx_i=(x_{i1}, \ldots, x_{ip})^{\small\mathrm{T}}\in\mathbb{R}^p$ is the feature vector of the $i$th observation, $y_i\in \{-1, +1\}$ is the class label, $n$ is the sample size, and $p$ is the number of predictors. Define $\boldsymbol{y} = [y_1,\ldots, y_n]^{\small\mathrm{T}}$ and $\vec{Y}$ as the $n \times n$ diagonal matrix with diagonal entries $\boldsymbol{y}$. 

A linear SVM classifier learns a decision function $f(\bx) = \bbetazero + \bbeta^{\small\mathrm{T}}\bx$, where $\bbeta =[\beta_1,\ldots, \beta_p]^{\small\mathrm{T}}$ and $\bbetazero \in \realset$, and classify any new observation $\bx$ by $\mbox{sign}[f(\bx)]$. Evaluating $f$ at all $n$ training points yields the vector $\boldsymbol{f}= [f(\bx_1),\ldots, f(\bx_n)]^{\small\mathrm{T}}$, which can be written compactly as $\boldsymbol{f}=\boldsymbol{e}\bbetazero + \vec{X}\bbeta$, where $\vec{X} =  [\bx_1,\ldots, \bx_n]^{\small\mathrm{T}}$ is the $n \times p$ design matrix and $\boldsymbol{e} = [1,\ldots, 1]^{\small\mathrm{T}}$ is the vector of ones. 

For the weighted SVM, the hinge loss of $f$ evaluated at the training set is $\boldsymbol{z} = [z_1,\ldots, z_n]^{\small\mathrm{T}}$, where $z_i = (1-y_if_i)_{+}$ for $i=1, \ldots, n$, and the weight vector is $\boldsymbol{W(y)} = [W(y_1),\ldots, W(y_n)]^{\small\mathrm{T}}$. For any input $\bx$, define the posterior class probabilities
$p(\bx)=P(Y=+1|\bX=\bx)$ and $1-p(\bx)=P(Y=-1|\bX=\bx)$. The optimal Bayes classifier signs $\bx$ to Class $+1$ if $p(\bx) \ge 0.5$, and to Class $-1$ otherwise. The same notation is used throughout Sections \ref{sec:l2wsvmmethod}-\ref{sec:elasticnetwsvm}.

\subsection{Weighted SVMs for Binary Classification}
\label{sec2:l2svmmethod}
In the weighted SVM (wSVM) framework, each class receives a different weight: $\pi\in[0,1]$ for Class $-1$; and $1-\pi$ for Class $+1$. The decision function $f$ is obtained by solving the following regularization problem
\begin{equation}
     \min_{f \in \mathcal{F}} ~~ \frac{1}{n}\Big[(1-\pi)\sum_{y_i=1}(1-{y_if(\bx_i)})_{+}+\pi\sum_{y_i=-1}(1-{y_if(\bx_i)})_{+}\Big]+\lambda J(f),
            \label{wsvml2}
\end{equation}
where $(1 - yf(\mathbf{x}))_+ = \max\{0, 1 - yf(\mathbf{x})\}$ denotes the hinge loss, $\mathcal{F}$ is some function space, and $J(f)$ is the penalty function. For kernel SVMs, we assume $\mathcal{F}$ to be
$\mathcal{H}_\mathbf{K}$, the reproducing kernel Hilbert space \citep[RKHS;][]{Wahba90} associated with a bivariate Mercer kernel $\mathbf{K}(\cdot,\cdot)$, and the penalty $J(f)=
\norm1[\big]f_{\mathcal{H}_\mathbf{K}}^2$, which is a regularization term that controls model complexity, with $\lambda>0$ governing the bias–variance trade-off \citep{Hastie2009}. By the representer theorem \citep{KW1971}, the minimizer of \eqref{wsvml2} admits the finite representation $f_{\pi}^{\lambda}(\mathbf{x}) = d + \sum_{i=1}^{n} c_i \mathbf{K}(\mathbf{x}, \mathbf{x}_i)$ for some coefficients $\boldsymbol{c} = [c_1, \ldots, c_n]^{\mathrm{T}} \in \mathbb{R}^n$ 
and $d\in\mathbb{R}$. The solution can be obtained through quadratic programming (QP), whose computational cost depends primarily on $n$ through duality \citep{lin_support_2002}.

For any weight $\pi$, the minimizer of the expected weighted hinge loss $\mathop{\mathbb{E}}_{(\bX,Y)\sim P(\bX, Y)}
\left\{W(Y)L[Yf(\bX)]\right\}$ has the same sign as 
$\mbox{sign}[p_{+1}(\bX)-\pi]$ \citep{WSL2008}. The class probability can then be estimated by training a series of classifiers $\hat{f}_{\pi_1},\cdots, \hat{f}_{\pi_{M}}$ via \eqref{wsvml2} over a grid of weights $0<\pi_1<\cdots<\pi_{M}<1$. Since $\hat{f}_\pi(\bx)$ is non-increasing in $\pi$ for any $\bx$, there exists a unique index 
$m^*$ such that $\hat{f}_{\pi_{m^*}}(\bx) = +1$ and $\hat{f}_{\pi_{m^*+1}}(\bx) = -1$, implying $\pi_{m^*}<p_{+1}(\bx)<\pi_{m^*+1}$. This yields a consistent probability estimator $\hat{p}_{+1}(\bx)=\frac{1}{2}(\pi_{m^*}+\pi_{m^*+1})$, whose precision is controlled by $M$, the density of the weight grid. This approach, based on the $L_2$-wSVM, is referred to as ``LTWSVM." The LTWSVM performs well on datasets with dense features
\citep{WSL2008,wang_multiclass_2019, zeng_wsvms_2022}, but becomes less efficient in the presence of redundant noise features \citep{TanWT10,ghaddar_high_2018}.

\section{Proposed Methodology: Two-Stage Linear wSVMs}\label{l1l2wsvm}

In this section, we introduce a sparse learning framework for weighted linear SVMs that enables automatic variable selection and probability estimation. We propose a two-stage approach: Stage 1 fits an $L_1$-wSVM to obtain a sparse solution and select relevant variables, and Stage 2 fits an $L_2$-wSVM to estimate class probabilities. Specifically, the $L_1$-wSVM in Stage 1 minimizes the weighted hinge loss for $f(\bx) = \bbetazero + \bbeta^{\small\mathrm{T}}\bx$, where $\bbetazero \in \realset$ and $\bbeta \in \realset^p$, subject to an $L_1$ penalty:
\begin{equation}
     \min_{\bbetazero,\bbeta}\enskip\frac{1}{n}\Big[(1-\pi)\sum_{y_i=1}(1-{y_if(\bx_i)})_{+}+\pi\sum_{y_i=-1}(1-{y_if(\bx_i)})_{+}\Big]+\lambda_1 \sum_{j=1}^{p} \Card{\beta_j}.
    \label{l1wsvms1}
\end{equation}
We adopt the same notation as in \ref{sec2:notations}, and $\lambda_1>0$ is the regularization parameter. 

To solve the problem \eqref{l1wsvms1}, we introduce the slack variables $u_j,v_j \ge 0$ for $j=1, \ldots, p$. Let $\beta_j = u_j-v_j$, The solution has $u_j$ or $v_j$ equal to 0, depending on the sign of $\beta_j$, we have $\Card{\beta_j} = u_j + v_j$. Define $\bbeta = \bu-\bv$ and the weight vector $\boldsymbol{W(y)} = [W(y_1),\ldots, W(y_n)]^{\small\mathrm{T}}$, where $W(y_i) = \pi$ if $y_i = -1$, and $(1- \pi)$ if $y_i = +1$ for $i=1, \ldots, n$. Now we can reformulate the unconstrained optimization problem in \eqref{l1wsvms1} as the constrained primal problem:
\begin{mini}[2]
  {\tiny{\beta_0,\bu,\bv,\bz}}{\boldsymbol{W(y)}^{\small\mathrm{T}}\boldsymbol{z} + n\lambda_1\sum_{j=1}^{p}(u_j + v_j) \label{optprimal1svm}}{}{}
  \addConstraint{\boldsymbol{z}}{\ge \boldsymbol{e} - \vec{Y}\vec{X}(\bu-\bv) - \vec{Y}\boldsymbol{e}\beta_0}, {}
\addConstraint{\boldsymbol{z},\boldsymbol{u},\boldsymbol{v}}{\ge 0.}{}{}
 \end{mini}
Introducing Lagrange multipliers $\boldsymbol{\gamma},\boldsymbol{\alpha},\boldsymbol{s},\boldsymbol{t} \ge 0$,  we obtain the Lagrange function of \eqref{optprimal1svm} as:
\begin{equation}  \begin{split} \label{lagfun1l1twsvm}
\max_{\boldsymbol{\alpha},\boldsymbol{\gamma},\boldsymbol{s},\boldsymbol{t}} \min_{\beta_0,\boldsymbol{z},\bu,\bv} \quad\mathcal{L} &= \boldsymbol{W(y)}^{\small\mathrm{T}}\boldsymbol{z} + n\lambda_1\boldsymbol{e}^{\small\mathrm{T}}(\bu+\bv) -  \boldsymbol{\gamma}^{\small\mathrm{T}}\boldsymbol{z} + \boldsymbol{e}^{\small\mathrm{T}}\boldsymbol{\alpha}  \\
    & \quad - \boldsymbol{\alpha}^{\small\mathrm{T}}\vec{Y}\vec{X}(\bu-\bv)-
    \boldsymbol{\alpha}^{\small\mathrm{T}}\vec{Y}\boldsymbol{e}\beta_0-
    \boldsymbol{\alpha}^{\small\mathrm{T}}\boldsymbol{z}-\bs^{\small\mathrm{T}}\bu - \bt^{\small\mathrm{T}}\bv. \
\end{split}
\end{equation}

\noindent
\red{To} minimize $\mathcal{L}$ with respect to $\boldsymbol{z},\bu,\bv, \beta_0$, by differentiating, we have:

\begin{enumerate}[labelwidth=1.5cm,labelindent=10pt,leftmargin=1.8cm,label=\bfseries Step \arabic*.,align=left]
 
\item[(1).] $\boldsymbol{\alpha}+\boldsymbol{\gamma}=\boldsymbol{W(y)}$ 
\item[(2).] $n\lambda_1\be - \vec{X}^{\small\mathrm{T}}\vec{Y}\boldsymbol{\alpha} -\bs =0$, since $\bs \ge 0$, we have $\vec{X}^{\small\mathrm{T}}\vec{Y}\boldsymbol{\alpha} \le n\lambda_1\be$
\item[(3).] $n\lambda_1\be + \vec{X}^{\small\mathrm{T}}\vec{Y}\boldsymbol{\alpha} -\bt =0$, since $\bt \ge 0$, we have $\vec{X}^{\small\mathrm{T}}\vec{Y}\boldsymbol{\alpha} \ge -n\lambda_1\be$
\item[(4).] $\boldsymbol{e}^{\small\mathrm{T}}\vec{Y}\boldsymbol{\alpha} = 0$

\end{enumerate}
By substituting above equations in \eqref{lagfun1l1twsvm}, \red{the} Wolfe dual of \eqref{optprimal1svm} reduces to solving 
\begin{mini}[2]
  {\boldsymbol{\alpha}}{-\be^{\small\mathrm{T}}\boldsymbol{\alpha} \label{optdual11svm}}{}{}
\addConstraint{\boldsymbol{y}^{\small\mathrm{T}}\boldsymbol{\alpha}}{=0}{}
  \addConstraint {0 \le \boldsymbol{\alpha}\le \boldsymbol{W(y)}}{}{}
\addConstraint {\vec{X}^{\small\mathrm{T}}\vec{Y}\boldsymbol{\alpha} \le n\lambda_1\be}{}
\addConstraint {\vec{X}^{\small\mathrm{T}}\vec{Y}\boldsymbol{\alpha} \ge -n\lambda_1\be.}{}
 \end{mini}

The dual problem \eqref{optdual11svm} is a standard linear program (LP) that can be solved using many programming platforms, including R, Python, and MATLAB. Since strong duality holds for LPs when a feasible solution exists \citep{boyd2004convex}, solving \eqref{optdual11svm} yields the same optimal solution as the primal problem \eqref{optprimal1svm}. Given the optimal $\boldsymbol{\alpha}^*$
from \eqref{optdual11svm}, we obtain $\bs = n\lambda_1\be - \vec{X}^{\small\mathrm{T}}\vec{Y}\boldsymbol{\alpha}^*$ and $\bt = n\lambda_1\be +\vec{X}^{\small\mathrm{T}}\vec{Y}\boldsymbol{\alpha}^*$
By the complementary slackness conditions of the Karush-Kuhn-Tucker (KKT) optimality conditions for LPs, we have $s_j u_j = 0$
and $t_j v_j = 0$ for $j \in \{1, \ldots, p\}$. When $s_j > 0$ and $t_j > 0$, it follows that $u_j = 0$ and $v_j = 0$, and hence $\beta_j = 0$, indicating that the $j$-th variable can be removed. This yields the selected variable set 
$\mathcal{V}^* = \{j : \beta_j \neq 0\}$. We also define the variable selection indicator, a $p$-vector $\mathcal{I}^*$, as $\mathcal{I}^*[j] = 1$ if $j\in\mathcal{V}^*$ and 
$\mathcal{I}^*[j] = 0$ otherwise.

At Stage 1, a sequence of weights $\pi_\epsilon \in \Pi_M = \{\frac{j-1}{m} \mid j=2, \ldots, m\}$ is applied to the $L_1$-wSVMs in \eqref{l1wsvms1}, where $m$ controls the precision of the estimated probabilities. The resulting subset of selected variables $\mathcal{V}^*_{\pi_\epsilon}$ is passed as input to the $L_2$-wSVMs with the corresponding weight for class probability estimation in Section \ref{sec2:l2svmmethod}. We refer to this two-stage procedure as ``LOTWSVM'' learning. Because the $L_1$-wSVM produces an important variable set $\mathcal{I}_\pi$ that may differ across weights, we aggregate the selected variables across all weights and compute the selection frequency as $\mathcal{F} = \sum_{\pi_\epsilon \in \Pi_M} \mathcal{I}_{\pi_\epsilon}^*$. The final set of important variables is identified by thresholding: $\mathcal{V}_{s} = \{j: \mathcal{F}[j] > s\}$, where $s$ is the frequency threshold for automatic variable selection.

The complete LOTWSVM algorithm is outlined in \textbf{Algorithm \ref{alg:algo1_lil2wsvm}}. The regularization parameters $\lambda_1 > 0$ in the $L_1$-wSVMs and $\lambda_2>0$ in the 
$L_2$-wSVMs are held fixed in Steps 2 and 3, respectively, but need to be tuned for optimal performance. We discuss parameter tuning in detail in Section \ref{sec:tuning}.

\begin{algorithm}
\caption{Sparse Learning with LOTWSVM} \label{alg:algo1_lil2wsvm}
\begin{algorithmic}[1] 

\State Initialize $\Pi_M = \{\frac{j-1}{m} \mid j=2, \ldots, m\}$.

\State For each $\pi_\epsilon \in \Pi_M$, fixed $\lambda_1 > 0$, train the $L_1$-wSVMs by minimizing \eqref{l1wsvms1}. Denote the selected variables as $\mathcal{V}^*_{\pi_\epsilon}$ and the variable indicator $\mathcal{I}_{\pi_\epsilon}^*$;

\State For each $\pi_\epsilon \in \Pi_M$ and fixed $\lambda_2 > 0$, use $\mathcal{V}^*_{\pi_\epsilon}$ as the new input, fit a linear $L_2$-wSVM by minimizing \eqref{wsvml2}, and obtain the estimated class probability $\hat{p}(\bx)$ based on \red{S}ection \ref{sec2:l2svmmethod};

\State For any data $\bx$, predict its class label as $\{+1\}$ if $\hat{p}(\bx) \ge 0.5$, and $\{-1\}$ otherwise. The decision boundary is given by $\hat{p}(\bx) = 0.5$;

\State Define the final set of important variables as $\mathcal{V}_{s} = \{j: \mathcal{F}[j] > s\}$, where
$\mathcal{F} = \sum_{\pi_\epsilon \in \Pi_M}$ is the selecton frequency and $s$ is the threshold.
\end{algorithmic}
\end{algorithm}

\section{Proposed ElasticNet wSVMs Learning Scheme}\label{sec:elasticnetwsvm}
\subsection{wSVMs with Elastic Net Penalty}
In the two-stage wSVMs, the $L_1$ penalty employed in Stage 1 lacks the grouping effect: when predictors are highly correlated, it tends to select only one variable from a correlated group while shrinking the rest to zero, potentially leading to numerical instability and misleading interpretations. Moreover, when $p>n$, the $L_1$ penalty can select at most $n$
variables before the objective function saturates, regardless of how many relevant variables exist. To address these limitations, we propose wSVMs with an elastic net penalty for class probability estimation and prove that this formulation achieves a grouping effect and can select more than $n$ important variables.

We \red{propose} two learning schemes to achieve grouping effects for variable selection, (1) fit the wSVM with an elastic penalty for automatic variable selection, follow\red{ed} by an $L_2$-wSVMs for class probability estimation; call this scheme ``ElasticNet-$L_2$ wSVMs"; (2) fit the wSVM with an elastic net penalty only for both variable selection and probability estimation; call this scheme ``ElasticNet wSVMs". To fit the wSVM with an elastic net penalty, we minimize the hinge loss of $f(\bx) = \bbetazero + \bbeta^{\small\mathrm{T}}\bx$ by solving
\begin{equation}
     \min_{\bbetazero,\bbeta}\enskip\frac{1}{n}\Big[(1-\pi)\sum_{y_i=1}(1-{y_if(\bx_i)})_{+}+\pi\sum_{y_i=-1}(1-{y_if(\bx_i)})_{+}\Big]+\lambda_1 \sum_{j=1}^{p} \Card{\beta_j} + \frac{\lambda_2}{2}\sum_{j=1}^{p}\beta_j^2,
    \label{elasticnetwsvms1}
\end{equation}
where $\lambda_1>0$ and $\lambda_2>0$ are the regularization parameters. We use the same notation \red{as} in \ref{sec2:notations}. First, we introduce the slack variables $\bu,\bv>0$ to relax the $\ell^1$-norm, which is the same as in Section \ref{l1l2wsvm}, and reformulate the unconstrained optimization problem in \eqref{elasticnetwsvms1} as a constrained primal problem
\begin{mini}[2]
  {\tiny{\beta_0,\bu,\bv,\bz}}{\boldsymbol{W(y)}^{\small\mathrm{T}}\boldsymbol{z} + n\lambda_1\sum_{j=1}^{p}(u_j + v_j) +\frac{n\lambda_2}{2}\sum_{j=1}^{p}(u_j + v_j)^2 \label{optprimalelasticneticwsvm}}{}{}
  \addConstraint{\boldsymbol{z}}{\ge \boldsymbol{e} - \vec{Y}\vec{X}(\bu-\bv) - \vec{Y}\boldsymbol{e}\beta_0}{}
  \addConstraint{\boldsymbol{z},\boldsymbol{u},\boldsymbol{v}}{\ge 0.}{}{}
 \end{mini}
Then, we introduce the Lagrange multipliers $\boldsymbol{\gamma},\boldsymbol{\alpha},\boldsymbol{s},\boldsymbol{t} \ge 0$ and obtain the Lagrange function \eqref{optprimalelasticneticwsvm}
\begin{equation}  \begin{split} \label{lagfun1elasticnetwsvm}
\max_{\boldsymbol{\alpha},\boldsymbol{\gamma},\boldsymbol{s},\boldsymbol{t}} \min_{\beta_0,\boldsymbol{z},\bu,\bv} \quad\mathcal{L} &= \boldsymbol{W(y)}^{\small\mathrm{T}}\boldsymbol{z} + n\lambda_1\boldsymbol{e}^{\small\mathrm{T}}(\bu+\bv) +
    \frac{n\lambda_2}{2}(\bu^{\small\mathrm{T}}\bu+\bv^{\small\mathrm{T}}\bv) -  \boldsymbol{\gamma}^{\small\mathrm{T}}\boldsymbol{z}\\
    & \quad  + \boldsymbol{\alpha}^{\small\mathrm{T}}\left[\boldsymbol{e}-\vec{Y}\vec{X}(\bu-\bv)-\vec{Y}\boldsymbol{e}\beta_0 -\bz\right]-\bs^{\small\mathrm{T}}\bu - \bt^{\small\mathrm{T}}\bv.  \
\end{split}
\end{equation}
By taking the derivative\red{s} with respect to $\boldsymbol{z},\bu,\bv, \beta_0$, we get
\begin{enumerate}[labelwidth=1.5cm,labelindent=10pt,leftmargin=1.8cm,label=\bfseries Step \arabic*.,align=left]

\item[(1).] $\boldsymbol{\alpha}+\boldsymbol{\gamma}=\boldsymbol{W(y)}$ 
\item[(2).] $\bu = \frac{1}{n\lambda_2}(\vec{X}^{\small\mathrm{T}}\vec{Y}\boldsymbol{\alpha} + \bs - n\lambda_1\be)$
\item[(3).] $\bv = \frac{1}{n\lambda_2}(-\vec{X}^{\small\mathrm{T}}\vec{Y}\boldsymbol{\alpha} + \bt - n\lambda_1\be)$
\item[(4).] $\boldsymbol{e}^{\small\mathrm{T}}\vec{Y}\boldsymbol{\alpha} = 0$
\end{enumerate}
Denote $\boldsymbol{e}_k$ as the vector of ones of length $k$, $\boldsymbol{0}_k$ as the vector of zeros of length $k$, $\boldsymbol{0}_{s,k}$ as the $s \times k$ zero matrix, and $\vec{I}_k$ as $k \cross k$ identity matrix. Define $\vec{A} = \left[\frac{1}{n\lambda_2}\vec{X}^{\small\mathrm{T}}\vec{Y} \enskip \frac{1}{n\lambda_2}\vec{I}_p \enskip \boldsymbol{0}_{p,p} \enskip -\frac{\lambda_1}{\lambda_2}\vec{I}_p\right]$ and $\vec{B} = \left[-\frac{1}{n\lambda_2}\vec{X}^{\small\mathrm{T}}\vec{Y} \enskip \boldsymbol{0}_{p,p} \enskip \frac{1}{n\lambda_2}\vec{I}_p \enskip   -\frac{\lambda_1}{\lambda_2}\vec{I}_p\right]$, both of size $p\times (n+3p)$, and define the vectors $\bc^{\small\mathrm{T}} = \left[\be_n^{\small\mathrm{T}} \enskip \boldsymbol{0}_{3p}^{\small\mathrm{T}}\right]$ and $\bw^{\small\mathrm{T}} =\left[\boldsymbol{\alpha}^{\small\mathrm{T}}\enskip \bs^{\small\mathrm{T}}\enskip \bt^{\small\mathrm{T}}\enskip \be_p^{\small\mathrm{T}}\right]$. Now we have $\bu = \vec{A}\bw$ and  $\bv = \vec{B}\bw$. Define $\vec{Q} = n\lambda_2 (\vec{A}^{\small\mathrm{T}}\vec{A}+\vec{B}^{\small\mathrm{T}}\vec{B}), \vec{D} = \left[\boldsymbol{0}_{p,n+2p}\enskip \vec{I}_p\right]$, and the $(2n+2p+1) \times (n+3p)$ constraint coefficients matrix $\vec{M}$ as:
$$\begin{pNiceMatrix}
   \boldsymbol{y}^{\small\mathrm{T}} & & \boldsymbol{0}_{1,3p}  \\
    \vec{I}_n & & \boldsymbol{0}_{n,3p} \\
    -\vec{I}_n & & \boldsymbol{0}_{n,3p} \\
    \boldsymbol{0}_{p,n} &  \vec{I}_p & & \boldsymbol{0}_{p,2p} \\
    \quad \boldsymbol{0}_{p,n+p} & & \vec{I}_p &  \boldsymbol{0}_{p,p} \\
\end{pNiceMatrix}.$$

The Wolfe dual of \eqref{optprimalelasticneticwsvm} is equivalent to solving the following optimization problem
\begin{mini}[2]
  {\bw}{\frac{1}{2}\bw^{\small\mathrm{T}}\vec{Q}\bw - \bc^{\small\mathrm{T}}\bw \label{optdualelasticnetwsvm1}}{}{}
\addConstraint{\vec{M}\bw\left[1\right]}{=0}{}
\addConstraint{\vec{M}\bw\left[2:2n+2p+1\right]}{\ge \boldsymbol{0}}{}
\addConstraint{\vec{D}\bw}{= \boldsymbol{e}_p}{}
\addConstraint{\vec{A}\bw}{ \ge \boldsymbol{0}}{}
\addConstraint{\vec{B}\bw}{ \ge \boldsymbol{0},}{}
\end{mini}
which is a standard quadratic programming (QP) problem. Applying \red{a} similar approach, the primal problem of \eqref{optprimalelasticneticwsvm} can be formulated as QP as well. 

Define $\bw^{\small\mathrm{T}} =\left[\bz^{\small\mathrm{T}}\enskip \bu^{\small\mathrm{T}}\enskip \bv^{\small\mathrm{T}}\enskip \bbetazero \right] $ and $\bd^{\small\mathrm{T}} =\left[\boldsymbol{W(y)}^{\small\mathrm{T}} \enskip n\lambda_1\boldsymbol{e}_{2p}^{\small\mathrm{T}}  \enskip 0 \right] $, both of length $(n+2p+1)$. And define the $(n+2p+1)$ square matrix $\vec{P}$ as
$$\begin{pNiceMatrix}
   & \quad \boldsymbol{0}_{n,n+2p+1} &  &  \\
    \enskip\boldsymbol{0}_{p,n} & \vec{I}_p & \boldsymbol{0}_{p,p+1} & \quad \\
    \enskip\quad \boldsymbol{0}_{p,n+p} & \quad\quad\quad \vec{I}_p & \boldsymbol{0}_p \quad\\
    & \quad \boldsymbol{0}_{n+2p+1}^{\small\mathrm{T}} & \\
\end{pNiceMatrix}.$$
Let the right hand side of the constraint coefficient vector be $\bb_0^{\small\mathrm{T}} =\left[\boldsymbol{0}_{n+2p}^{\small\mathrm{T}} \enskip \be_n^{\small\mathrm{T}}\right]$, which is of length $(2n+2p)$, and let the left hand side be the $(2n+2p) \cross (n+2p+1)$ constraint coefficients matrix $\vec{R}$
$$\begin{pNiceMatrix}
   \vec{I}_n & & \quad\enskip \boldsymbol{0}_{n,2p+1} & \\
    \enskip\boldsymbol{0}_{p,n} & \vec{I}_p & \quad \boldsymbol{0}_{p,p+1} & \\
    \enskip\quad \boldsymbol{0}_{p,n+p} & & \vec{I}_p & \boldsymbol{0}_p \\
    \vec{I}_n &  \vec{Y}\vec{X} & -\vec{Y}\vec{X} & \vec{Y}\be_n\\
\end{pNiceMatrix}.$$
Now we have the QP of primal problem \eqref{optprimalelasticneticwsvm} as:
\begin{mini}[2]
  {\bw}{\frac{1}{2}\bw^{\small\mathrm{T}}\vec{P}\bw + \bd^{\small\mathrm{T}}\bw \label{optprimalelasticnetwsvm2}}{}{}
\addConstraint{\vec{R}\bw}{ \ge \bb_0.}{}
\end{mini}
Since the strong duality holds for SVMs \citep{Eladio5400391,dekel_support_2016,WANG2022382}, \red{the} primal problem and \red{the} dual problem have same optimal solution. The QP for the primal and dual problems of ElasticNet wSVMs have different complexity and will \red{be discussed in S}ection \ref{sec:complex}. For the primal problem, we can obain the optimal $\bu^*$ and $\bv^*$ and $\bbetazero^*$ directly from \eqref{optprimalelasticnetwsvm2}, and $\hat{\bbeta}^{p} = \bu^*-\bv^*$, $\hat{\bbetazero}^{p} = \bbetazero^*$. For dual problem \eqref{optdualelasticnetwsvm1}, we obtain $\bu^* =\vec{A}\bw^* $, $\bv^* =\vec{B}\bw^*$, and $\boldsymbol{\alpha}^*$. We have $\hat{\bbeta}^{d} = \bu^*-\bv^*$. Based on the complementary slackness of Karush-Kuhn-Tucker (KKT) optimality condition, we have:\\
(1) $\alpha_i^*(1-y_i\hat{f}_i-z_i)=0$; \\
(2) $\gamma_i^*z_i=0$, $\forall i \in \{1,\ldots, n\}$, and $\{\gamma_i^* + \alpha_i^* = W(y_i), \gamma_i^*, \alpha_i^* \ge 0\}$. \\
For some $i \in \{1,\ldots,n\}$, we have $0 < \alpha_i^* < W(y_i)$, leading to $z_i =0$, then $1-y_i\hat{f}_i = 0$. Since $\hat{f}_i = \hat{f}(\bx_i) =\hat{\bbetazero}^{d} + \hat{\bbeta}^{d}\bx_i$, we have $ \hat{\bbetazero}^{d} = y_i^{-1}-\hat{\bbeta}^{d}\bx_i$. All the points with $\alpha_i^*$ satisfying \red{the} above condition are the support vectors of wSVMs, which will give the same $\hat{\bbetazero}^{d}$ value. To enahance numerical stability, we provide a robust solution
\begin{equation}
    \hat{\bbetazero}^{d}=\frac{\boldsymbol{e}^{\small\mathrm{T}}[\vec{I}\odot\boldsymbol{\alpha}^*][\vec{I}\odot(\boldsymbol{W(y)}-\boldsymbol{\alpha}^*][\boldsymbol{y}^{-1} - \vec{X}\hat{\bbeta}^d]}    
    {\boldsymbol{\alpha}^{*{\small\mathrm{T}}}(\boldsymbol{W(y)}-\boldsymbol{\alpha}^*)},
\end{equation}
where \vec{I} is the $n \times n$ identity matrix and $\odot$ is Hadamard product with \emph{broadcasting}. For a given weight $\pi$ and fixed tuning parameters $\lambda_1>0$ and $\lambda_2>0$, we denote the optimal classifier for \eqref{elasticnetwsvms1} as $\hat{f}_{\pi}^{\lambda_1,\lambda_2}(\bx) =  \hat{\bbetazero} + \hat\bbeta^{\small\mathrm{T}}\bx$, where $\hat{\bbetazero}$ and $\hat{\bbeta}$ are estimated by the primal or dual problem, and should provide the same values. For each $\pi$, the selected variable set $\mathcal{V}_{\pi}^* = \{j: \beta_{j} \neq 0 \}$, $j \in \{1,\ldots,p\}$, and the variable selection indicator $\mathcal{I}_{\pi}^*$, where $\mathcal{I}_{\pi}^*[j] = 1$ if $j \in \mathcal{V}_{\pi}^*$, and $\mathcal{I}_{\pi}^*[j] = 0$ otherwise. We discuss the parameter tuning in Section \ref{sec:tuning}.

\subsection{Refitting with $L_2$ wSVMs}
Fitting a weighted SVM with the elastic net penalty yields initial estimates of $\bbeta$ that are sparse due to the $\ell^1$-norm, which performs continuous variable selection. As with LOTWSVM, the selected variables can then be passed to $L_2$-wSVMs for class probability estimation. This combination may be advantagous for data with complex non-linear distributions, since the downstream $L_2$-wSVMs can employ non-linear kernels. Furthermore, the $\ell^2$-norm provides a grouping effect, allowing the number of selected variables to exceed $n$. Wed denote this 
procedure as ``ENTPWSVM'' when the sparse solution is obtained via the primal QP, or ``ENTWSVM" when obtained vias the dual QP. The complete algorithm is presented in Algorithm \ref{alg:algo2_en2wsvm}.

\begin{algorithm}
\caption{Sparse Learning with ENTPWSVM (primal) or ENTWSVM (dual)} \label{alg:algo2_en2wsvm}
\begin{algorithmic}[1] 

\State Initialize $\Pi_M = \{\frac{j-1}{m} \mid j=2, \ldots, m\}$;

\State For each $\pi_\epsilon \in \Pi_M$ and fixed $\lambda_1, \lambda_2 > 0$, train the ElasticNet wSVMs by minimizing \eqref{elasticnetwsvms1}, by solving either the primal QP \eqref{optprimalelasticnetwsvm2} as ``ENTPWSVM", or the dual QP \eqref{optdualelasticnetwsvm1} as ``ENTWSVM".  We obtain the sparse solution $\hat{\bbeta}$, the set of selected variables $\mathcal{V}^*_{\pi_\epsilon}$, and the variable indicator $\mathcal{I}_{\pi_\epsilon}^*$;

\State For each $\pi_\epsilon \in \Pi_M$ andfixed $\lambda_3 > 0$, apply the linear $L_2$-wSVM with reduced $\mathcal{V}^*_{\pi_\epsilon}$ by minimizing \eqref{wsvml2} and estimate the class probability $\hat{p}(\bx)$ based on Section \ref{sec2:l2svmmethod};

\State For any data $\bx$, predict its class label as $+1$ if $\hat{p}(\bx) \ge 0.5$, and $-1$ otherwise. The decision boundary is given by $\hat{p}(\bx) = 0.5$;

\State Define the final set of important variables as $\mathcal{V}_{s} = \{j: \mathcal{F}[j] > s\}$, where
$\mathcal{F} = \sum_{\pi_\epsilon \in \Pi_M}$ is the selecton frequency and $s$ is the threshold.

\end{algorithmic}
\end{algorithm}

\subsection{Learning with ElasticNet wSVMs}

Alternatively, the elastic net wSVMs can serve as a one-step approach that performs both automatic variable selection and class probability estimation directly, without involving the $L_2$-wSVMs. This formulation retains the grouping effect. We denote this procedure as
``ENPWSVM'' when the sparse solution is obtained via the primal QP, or ENWSVM'' when obtained via the dual QP. The complete algorithm is presented in Algorithm \ref{alg:algo3_enwsvm}.
\begin{algorithm}[H]
\caption{Sparse Learning with ENPWSVM (primal) or ENWSVM (dual)} \label{alg:algo3_enwsvm}
\begin{algorithmic}[1] 

\State Initialize $\Pi_M = \{\frac{j-1}{m} \mid j=2, \ldots, m\}$;

\State For each $\pi_\epsilon \in \Pi_M$ and fixed $\lambda_1, \lambda_2$, train the ElasticNet wSVMs by \eqref{elasticnetwsvms1}, through either the primal QP \eqref{optprimalelasticnetwsvm2} as ``ENPWSVM", or the dual QP \eqref{optdualelasticnetwsvm1} as ``ENWSVM".  We get the classifier $\hat{f}_{\pi_\epsilon}^{\lambda_1,\lambda_2}(\bx) =  \hat{\bbetazero} + \hat\bbeta^{\small\mathrm{T}}\bx$, the selected variables $\mathcal{V}^*_{\pi_\epsilon}$, and \red{the} variable indicator $\mathcal{I}_{\pi_\epsilon}^*$, based on the sparsity of $\hat\bbeta$; 

\State Replace the multiple classifiers $\hat{f}_{\pi_\epsilon}$ for $L_2$-wSVMs with  $\hat{f}_{\pi_\epsilon}^{\lambda_1,\lambda_2}(\bx)$ from step (2), and follow the procedure for class probability estimation in \ref{sec2:l2svmmethod} with predicted labels associated weights;

\State Predict class label $+1$ for $\bx$ if $\hat{p}(\bx) \ge 0.5$, and $-1$ otherwise. The decision boundary is $\hat{p}(\bx) = 0.5$;

\State Define the final set of important variables as $\mathcal{V}_{s} = \{j: \mathcal{F}[j] > s\}$, where
$\mathcal{F} = \sum_{\pi_\epsilon \in \Pi_M}$ is the selecton frequency and $s$ is the threshold.
\end{algorithmic}
\end{algorithm}  


\subsection{Theory: Grouping Effect of ElasticNet wSVMs}\label{variablegrouping}

Compared with the $L_1$-wSVMs, the elastic net wSVMs offer the advantage of variable grouping during selection. In this section, we prove this grouping effect for highly correlated variables. The result is general and holds for all Lipschitz continuous loss functions.

\begin{theorem}\label{thm1}
Denote the solution to \eqref{elasticnetwsvms1} as $\hat{\bbetazero}$ and $\hat{\bbeta}$. Assume that the loss function $L(\cdot)$ is Lipschitz continuous, i.e. $\Card{L(t_1) - L(t_2)} \le M\Card{t_1-t_2}$ for some positive infinite real constant $M$, then for any pair $(s, t)$, where $1 \le s < t \le p$, we have: 
\begin{equation}
    \Card{\hat{\beta_s} - \hat{\beta_t}} \le \frac{M}{n\lambda_2}\Big[(1-\pi)\sum_{y_i=1}\Card{x_{is}-x_{it}} +\pi\sum_{y_i=-1}\Card{x_{is}-x_{it}}\Big]. \label{thm1p}
\end{equation}
\end{theorem}

The proof is a straightforward extension of Theorem 1 in \cite*{wang_doubly_2006} and is therefore omitted. Since the hinge loss satisfies Lipschitz continuity with $M=1$, Theorem 1 holds for the elastic net wSVMs. The grouping effect depends on $\lambda_2$ from the $\ell^2$-norm penalty in the elastic net.

\section{Complexity Analysis}\label{sec:complex}
In this section, we provide a complexity analysis of the proposed learning methods: the $L_2$-wSVMs (LTWSVM) and the two-stage wSVMs (LOTWSVM), which are solved via their dual formulations; and the elastic net wSVMs (ENPWSVM/ENWSVM) and elastic net-$L_2$ wSVMs (ENTPWSVM/ENTWSVM), which can be solved via either the primal or dual formulation.

We adopt the same notation without loss of generality. Assume the binary classification problem is balanced, i.e., $n_{+1} = n_{-1}$, and that the data are randomly split into training and tuning sets of sizes $n_{train}$
and $n_{tune}$, respectively, preserving the class balance. The data dimension is $p$. For the convex optimization problems derived from the wSVMs with hinge loss, let $N$ denote the number of variables in the objective and $D$ the number of constraints. 

We solve quadratic programs (QP) in R using the \texttt{quadprog} package, which employs an active set algorithm with polynomial time complexity $C_{QP} \sim O(\max(N^3, N^2D))$
\citep{Tseng1988ASP,ye_extension_1989,ye_complexity_1998,boyd2004convex}. For linear programs (LP), we use \texttt{lpSolveAPI}, an R interface to the \texttt{lp\_solve} mixed integer linear programming solver \citep{Berkelaar2004}, with complexity $C_{LP} \sim O(\max(N^{2.5}, ND^{1.5}))$ \citep{Vaidya63499}. The time complexity for label prediction is $C_{LLT} \sim O(n_{tune} \, n_{train})$ for the $L_2$ kernel wSVMs and $C_{LEN} \sim O(n_{tune})$ for the elastic net wSVMs. Training a sequence of $O(m)$ wSVMs for binary class probability estimation on $n_{tune}$
data points has complexity $C_{P} \sim n_{tune} \, O(m \log m)$. For parameter tuning, let $n_{par}$ denote the number of tuning parameters and $\{\gamma_p \mid p \in \{1, \ldots, n_{par}\}\}$ is the corresponding number of grid points. Table \ref{table1:complexity} summarizes the time complexity of the proposed sparse learning methods.

\begin{table}[H]
\caption{Complexity Analysis of Sparse Learning with wSVMs.}
\centering
\scalebox{0.79}{
\centering
\begin{tabular}{c|c|c|c|c|c} 
\hline
\textbf{Methods} & \textbf{Opt} & \textbf{$N$} & \textbf{$D$} & \textbf{Solver} & \textbf{Complexity}  \\ 
\hline
LTWSVM           & Dual                  & $n$          & $2n+1$       & \texttt{quadprog}       & $(\prod_{p=1}^{n_{par}}\gamma_p)\left[O(m)(C_{QP}  + C_{LLT}) + C_{P}\right]      $              \\
\hline
LOTWSVM          & Dual                  & $n$          & $2n+2p+1$    & \texttt{lpSolveAPI}      & $(\prod_{p=1}^{n_{par}}\gamma_p)\left[O(m)(\max(C_{QP},C_{LP})  + C_{LLT}) + C_{P}\right]  $                   \\
\hline
ENPWSVM          & Primal                & $n+2p+1$     & $2n+2p$      & \texttt{quadprog}       & $(\prod_{p=1}^{n_{par}}\gamma_p)\left[O(m)(C_{QP} + C_{LEN}) + C_{P}\right]  $                  \\
ENWSVM           & Dual                  & $n+3p$       & $2n+5p+1$    & \texttt{quadprog}        & $(\prod_{p=1}^{n_{par}}\gamma_p)\left[O(m)(C_{QP} + C_{LEN}) + C_{P}\right]$     \\
\hline
ENTPWSVM          & Primal                & $n+2p+1$     & $2n+2p$      & \texttt{quadprog}       & $(\prod_{p=1}^{n_{par}}\gamma_p)\left[O(m)(C_{QP} + C_{LLT}) + C_{P}\right]    $                \\
ENTWSVM           & Dual                  & $n+3p$       & $2n+5p+1$    & \texttt{quadprog}        & $(\prod_{p=1}^{n_{par}}\gamma_p)\left[O(m)(C_{QP} + C_{LLT}) + C_{P}\right]      $       \\
\hline
\end{tabular}}
\caption*{\footnotesize
    NOTE:  The table lists the time complexity for the proposed sparse learning methods based on wSVMs. Opt: optimization method; $N$: the number of variables in the objective; $D$: the number of constraints; $n$: the number of training observations; $p$: data dimensionality. LOTWSVM uses the \texttt{quadprog} solver for probability estimation. }
\label{table1:complexity}
\end{table}

We now illustrate the complexity of each method using linear kernel wSVMs. The LTWSVM has a single hyperparameter $\lambda>0$ for the $L_2$-wSVMs; the LOTWSVM has two hyperparameters, $\lambda_1 > 0$ for the $L_1$-wSVMs and $\lambda_2 > 0$ for the $L_2$-wSVMs; the ENPWSVM/ENWSVM has two hyperparameters $\lambda_1, \lambda_2 > 0$ for the elastic net wSVMs; and the ENTPWSVM/ENTWSVM has three hyperparameters, $
\lambda_1, \lambda_2 > 0$ for the elastic net wSVM and $\lambda_3 > 0$
for the $L_2$-wSVM. For each hyperparameter, we set the number of grid points to $\gamma_\lambda = \tau$. Assume fixed precision with $m$
constant and $n_{train} = n_{tune} = n$. To reduce tuning complexity, we set $\lambda_1 = \lambda_2$ in LOTWSVM and set $\lambda_2 = \lambda_3$
in ENTPWSVM/ENTWSVM, as the additional tuning loop yields no significant performance improvement. The complexities from Table 1 then reduce to: (1) LTWSVM: O$(\tau n^{3})$; (2) LOTWSVM: O$(\tau \max(n^3, np^{1.5}))$ ; (3) ENPWSVM/ENTPWSVM: O($\tau^2 (n+2p)^{3}$); and (4) ENWSVM/ENTWSVM: O($\tau^2 (n+3p)^{3}$). In the low-dimensional case ($p \ll n$), all methods share the same per-loop complexity of $O(n^3)$; in the high-dimensional case ($p \gg n$), the complexity of the sparse learning methods grows with $p$, with the elastic net wSVMs dual problem being the most expensive.

Solving QPs and LPs with \texttt{quadprog} and \texttt{lp\_solve} is computationally expensive for high-dimensional data. To improve efficiency, we adopt the \texttt{OSQP} solver, which uses a modified alternating direction method of multipliers (ADMM) algorithm \citep{Boyd220016} and is typically an order of magnitude faster than \texttt{quadprog} \citep{Stellato_osqp_20}. The \texttt{OSQP} solver requires the symmetric matrix $Q$ in the QP objective to be positive semidefinite. We adapt it to handle LPs by setting $Q = \boldsymbol{0}_{N,N}$. In Section \ref{sec:simu}, we use Example 1 to compare the empirical complexity of all methods across different solvers, confirming the theoretical analysis.

\section{Parameter Tuning}\label{sec:tuning}
The regularization parameters, denoted by $\Lambda > 0$, must be tuned adaptively from the data, as their values are critical to the performance of the wSVMs. For LTWSVM and LOTWSVM, $\Lambda = \{\lambda\}$, and for the elastic net wSVMs, $\Lambda = \{\lambda_1, \lambda_2\}$. To select appropriate values of $\Lambda$, we randomly split the data into two parts of appropriately equal sample sizes: a training set $\mathbb{S}_{train}$ used to fit the wSVM classifiers for a given weight $\pi$ and fixed $\Lambda$, and a tuning set $\mathbb{S}_{tune}$ used to choose the optimal values of $\Lambda$.

For the one-step approaches (LTWSVM and ENWSVM/ENPWSVM), given a fixed $\Lambda$, we train $m-1$ classifiers $\hat{f}_{\pi_{\epsilon}}^\Lambda(\bx)$ for 
$\pi_\epsilon \in \{\frac{j-1}{m} \mid j=2, \ldots, m\}$ on $\mathbb{S}_{train}$ and estimate the probability of $\hat{p}^{\Lambda}$ on $\mathbb{S}_{tune}$. For the two-step approaches (LOTWSVM and ENTWSVM/ENTPWSVM), we proceed as follows. First, for a fixed 
$\Lambda_1 \in \Lambda$, where $\Lambda_1 = \lambda$ for LOTWSVM and 
$\Lambda_1 = \{\lambda_1, \lambda_2\}$ for ENTWSVM/ENTPWSVM, we train 
$m-1$ classifiers $\hat{f}_{\pi_{\epsilon}}^{\Lambda_1}(\bx)$ on 
$\mathbb{S}_{train}$ to perform automatic variable selection, reducing the data to $\bx'$ with corresponding subsets $\mathbb{S}_{train}^{{\Lambda_1}'}$ and $\mathbb{S}_{tune}^{{\Lambda_1}'}$.
Second, for a fixed $\Lambda_2 \in \Lambda$, where $\Lambda_2 = \lambda$
for LOTWSVM and $\Lambda_2 = \lambda_2$ for ENTWSVM/ENTPWSVM, we train 
$m-1$ classifiers $\hat{f}_{\pi_{\epsilon}}^{\Lambda_2}(\bx')$ on
$\mathbb{S}_{train}^{{\Lambda_1}'}$ to estimate $\hat{p}^{\Lambda}$
on $\mathbb{S}_{tune}^{{\Lambda_1}'}$. The tuning parameters are selected via grid search based on the performance of $\hat{p}^{\Lambda}$ on the tuning set. Let $p(\bx)$ denote the true class probability $P(Y=+1|\bx)$. To quantify the estimation accuracy of $\hat{p}$, we use the generalized Kullback-Leibler (GKL) loss function:
\begin{eqnarray}    
&&GKL(p, \hat {p}^{\Lambda})\label{eq:gkl1}\\
&=&\mathlarger{\mathop{\mathbb{E}}}_{\bX\sim \mathcal{P}}\left[p(\bX)\log \frac{p(\bX)}{\hat p^{\Lambda}(\bX)}+(1-p(\bX))\log \frac{1-p(\bX)}{1-\hat p^{\Lambda}(\bX)}\right]\nonumber    \\
&=&\mathlarger{\mathop{\mathbb{E}}}_{\bX\sim \mathcal{P}}\left[p(\bX)\log {p(\bX)}+(1-p(\bX))\log( {1-p(\bX)})\right]\label{eq:gkl2}    \\ 
&\;&- \mathlarger{\mathop{\mathbb{E}}}_{\bX\sim \mathcal{P}}\left[p(\bX)\log {\hat p^{\Lambda}(\bX)}+(1-p(\bX))\log( {1-\hat p^{\Lambda}(\bX)})\right].\label{eq:gkl3}
\end{eqnarray}
The expectation is taken with respect to the distribution $\mathcal{P}$ of
$\bX$. The term \eqref{eq:gkl2} does not depend on $\hat{p}^{\Lambda}(\bX)$
and can be treated as a constant, so only 
\eqref{eq:gkl3} is relevant. In practice, the true class probability $p(\bX)$ is generally unknown, so we develop an empirical estimator to approximate GKL based on the tuning data. Noting that $\mathlarger{\mathop{\mathbb{E}}}\left[\frac{1}{2}(Y+1) \mid \bX, Y \in \{+1,-1\}\right] = p(\bX)$, we can approximate the GKL by
\begin{equation}
    EGKL(\hat p^{\Lambda})\label{eq:egkl} = -\frac{1}{2n}\sum_{i=1}^n\left[(1+y_i)\log \hat p^{\Lambda}(\bx_i)
+ (1-y_i) \log (1-\hat p^{\Lambda}(\bx_i))\right]
\end{equation}    
The optimal values of $\Lambda$ are chosen by minimizing the EGKL.

\section{Numerical Studies}\label{sec:simu}
In this section, we evaluate the proposed sparse learning methods based on two-stage $L_1$-wSVMs and elastic net wSVMs on simulated high-dimensional data, comparing them with $L_2$-wSVMs, which lack variable selection capability. We demonstrate that our methods effectively remove irrelevant variables, yielding improved class probability estimation and classification accuracy. We further show that the elastic net wSVMs identify highly correlated variables jointly through the grouping effect, achieving the best overall performance. All experiments were conducted on the University of Arizona High-Performance Computing (HPC) cluster using AMD Zen2 48-core processors with 15GB of memory.

For class probability estimation, we set the weights as $\pi_{\epsilon} = \frac{j-1}{m}, j = \{1, \ldots, m+1\}$, where $m = \floor{\sqrt{n}}$ controls the estimation precision and $n$ is the training set size. A linear kernel is used in the simulations. The tuning parameters 
$\lambda_1, \lambda_2\in \{5.5 \times 10^{-4}, 5.5 \times 10^{-3}, \ldots, 5.5 \times 10^{3}\}$ are selected by grid search using the EGKL criterion, which has been shown to approximate the GKL well \citep{zeng_wsvms_2022}. We set $\Card{\mathbb{S}_{train}} = \Card{\mathbb{S}_{tune}} = n$ and evaluate probability estimation accuracy on a test set 
$\mathbb{S}_{test} = \{(\tilde{\bx}_i, \tilde{y}_i), i = 1, \ldots, \tilde{n}\}$ with $\tilde{n} = 50n$, using the EGKL as the evaluation criterion. Classification follows the $\argmax$ rule, assigning $\bx$
to Class $+1$ when $\hat{p}(\bx) \ge 0.5$.

The proposed two-stage $L_1$-wSVMs and elastic net wSVMs perform automatic variable selection for each weight $\pi_{\epsilon}$. We introduce a threshold $P_{\beta}$: any coefficient with $|\beta_j| < P_{\beta}, j \in \{1, \ldots, p\}$ is set to zero, indicating the $j$th variable is irrelevant. The variable selection indicator for weight $\pi_{\epsilon}$ is the $p$-vector $\mathcal{I}_{\pi_{\epsilon}}^*$, where $\mathcal{I}_{\pi_{\epsilon}}^*[j] = 1$ if $|\beta_j| > P_{\beta}$ and
$\mathcal{I}_{\pi_{\epsilon}}^*[j] = 0$ otherwise. Over $N$ Monte Carlo simulations with $m-1$ wSVMs each, we aggregate the selected variables as $\mathcal{A} = \sum_{N} \sum_{\pi_\epsilon \in \Pi_M} \mathcal{I}_{\pi_\epsilon}^*$ and define the selection frequency as $\mathcal{F} = \mathcal{A}/N \in [0, m-1]$. The final variable set is determined by thresholding: $\mathcal{V}_{s} = \{j : \mathcal{F}[j] > s\}
$. In the following examples, we set $s = (m-1)/2$.

We consider three examples, each with five important variables at varying levels of correlation. We set $n = 100$ and $\tilde{n} = 5{,}000$
across four dimensionalities, $p \in \{100, 200, 400, 1000\}$. In Example 1, we first run $N = 50$ Monte Carlo simulations across all methods with different optimization approaches and solvers to assess computational cost and identify the best configuration for each method. We then run $N = 100$
Monte Carlo simulations for each example and report average performance with standard error se= $\sigma_{\bv}/\sqrt{N}$.

\medskip
\noindent
\textbf{Example 1 (Balanced, Independent Setting)}. In this example, the input variables are independent. Let $\boldsymbol{c}_k$ denote a $k$
-vector with all elements equal to $c$. Data are generated from a multivariate normal distribution with $\boldsymbol{\mu_{+1}} = [\boldsymbol{0.5}_{5} \enskip \boldsymbol{0}_{p-5}]^{\small\mathrm{T}}, \boldsymbol{\mu_{-1}} = [\boldsymbol{-0.5}_{5} \enskip \boldsymbol{0}_{p-5}]^{\small\mathrm{T}}$, and $\bSigma = \mathbf{I}_p$. The Bayes classifier depends only on $x_1, \ldots, x_5$, with a Bayes error of 0.132, independent of $p$.

We use this example to examine the computational cost of the proposed methods alongside the $L_2$-wSVMs (LTWSVM) across different optimization approaches, solvers, and dimensions $p$. Table \ref{table2:comparison} summarizes the results, which are consistent with the complexity analysis in Section \ref{sec:complex}. For all sparse learning methods, computation time increases with $p$, whereas LTWSVM depends only on the training sample size $n$. The primal and dual formulations yield identical results for the elastic net approaches, but solving the primal problem is over 20 times faster with the standard \texttt{quadprog} solver. Switching to \texttt{OSQP} provides an additional 15-fold speedup, making the primal solution roughly 300 times faster than the dual. The LOTWSVM has lower-order complexity than the elastic net approaches; with \texttt{OSQP}, it is over 10 times faster on high-dimensional data while achieving comparable performance to \texttt{lpSolveAPI}. Since \texttt{OSQP} offers no advantage over \texttt{quadprog} for LTWSVM, we use \texttt{quadprog} for LTWSVM and \texttt{OSQP} for LOTWSVM, ENPWSVM, and ENTPWSVM in all subsequent experiments.

\begin{table}[H]
\caption{Performance measure on wSVMs learning methods}
\centering
\resizebox{1\textwidth}{!}{
\begin{tabular}{c|cc|ccc|ccc|ccc|ccc} 
\hline
\multirow{2}{*}{\textbf{Methods}} & \multirow{2}{*}{\textbf{Opt}} & \multirow{2}{*}{\textbf{Solver}} & \multicolumn{3}{c|}{\textbf{n = 100, p =100}} & \multicolumn{3}{c|}{\textbf{n = 100, p =200}} & \multicolumn{3}{c|}{\textbf{n = 100, p =400}} & \multicolumn{3}{c}{\textbf{n = 100, p =1000}}  \\ 
\cline{4-15}
                                  &                               &                                  & \textbf{Time} & \textbf{EGKL} & \textbf{TE}   & \textbf{Time} & \textbf{EGKL} & \textbf{TE}   & \textbf{Time} & \textbf{EGKL} & \textbf{TE}   & \textbf{Time} & \textbf{EGKL} & \textbf{TE}    \\ 
\hline
LTWSVM                            & Dual                          & quadprog                         & 0.3 (0.0)     & 49.8 (0.5)    & 22.8 (0.3)    & 0.3 (0.0)     & 54.6 (0.3)    & 26.2 (0.2)    & 0.3 (0.0)     & 59.2 (0.3)    & 30.4 (0.2)    & 0.4 (0.1)     & 64.8 (0.2)    & 36.3 (0.3)     \\
LTWSVM                            & Dual                          & OSQP                             & 0.3 (0.0)     & 49.8 (0.5)    & 22.8 (0.3)    & 0.3 (0.0)     & 54.6 (0.3)    & 26.2 (0.2)    & 0.3 (0.0)     & 59.2 (0.3)    & 30.4 (0.2)    & 0.4 (0.1)     & 64.7 (0.2)    & 36.2 (0.2)     \\ 
\hline
LOTWSVM                           & Dual                          & lpSolveAPI                       & 0.5 (0.0)     & 42.2 (0.5)    & 17.7 (0.3)    & 0.7 (0.1)     & 43.8 (0.5)    & 17.9 (0.3)    & 1.6 (0.1)     & 44.4 (0.4)    & 18.8 (0.3)    & 17.8 (0.2)    & 47.6 (0.5)    & 20.5 (0.4)     \\
LOTWSVM                           & Dual                          & OSQP                             & 0.4 (0.0)     & 44.3 (0.6)    & 18.3 (0.4)    & 0.4 (0.0)     & 46.3 (0.6)    & 18.9 (0.3)    & 0.6 (0.0)     & 47.7 (0.6)    & 20.0 (0.3)    & 1.7 (0.1)     & 49.1 (0.5)    & 21.3 (0.3)     \\ 
\hline
ENPWSVM                           & Primal                        & quadprog                         & 1.1 (0.0)     & 44.7 (0.7)    & 17.3 (0.4)    & 2.8 (0.0)     & 45.5 (0.8)    & 18.0 (0.4)    & 11.3 (0.1)    & 44.4 (0.5)    & 18.6 (0.3)    & 153.9 (2.8)   & 46.7 (0.5)    & 20.9 (0.3)     \\
ENPWSVM                           & Primal                        & OSQP                             & 0.7 (0.0)     & 44.7 (0.7)    & 17.3 (0.4)    & 1.3 (0.0)     & 45.3 (0.7)    & 18.1 (0.5)    & 3.1 (0.0)     & 44.8 (0.7)    & 18.9 (0.4)    & 11.2 (0.3)    & 47.2 (0.7)    & 20.9 (0.4)     \\ 
\hline
ENWSVM                            & Dual                          & quadprog                         & 3.9 (0.0)     & 45.3 (0.7)    & 17.5 (0.4)    & 20.1 (0.1)    & 45.1 (0.6)    & 17.9 (0.4)    & 158.2 (3.1)   & 44.5 (0.5)    & 18.7 (0.3)    & 2832.0 (34.7) & 46.8 (0.5)    & 20.9 (0.3)     \\
ENWSVM                            & Dual                          & OSQP                             & 1.0 (0.0)     & 45.4 (0.8)    & 17.4 (0.4)    & 1.9 (0.0)     & 45.9 (0.9)    & 18.2 (0.5)    & 4.5 (0.1)     & 44.9 (0.7)    & 18.9 (0.4)    & 16.7 (0.2)    & 46.9 (0.5)    & 21.0 (0.4)     \\ 
\hline
ENTPWSVM                          & Primal                        & quadprog                         & 1.1 (0.0)     & 43.5 (0.7)    & 17.6 (0.4)    & 2.8 (0.0)     & 45.0 (0.6)    & 18.2 (0.3)    & 11.2 (0.1)    & 46.4 (0.4)    & 19.1 (0.3)    & 151.7 (2.7)   & 48.7 (0.4)    & 20.5 (0.4)     \\
ENTPWSVM                          & Primal                        & OSQP                             & 0.8 (0.0)     & 43.2 (0.5)    & 18.3 (0.4)    & 1.4 (0.0)     & 44.9 (0.4)    & 18.1 (0.4)    & 3.1 (0.0)     & 47.1 (0.5)    & 18.4 (0.5)    & 11.2 (0.6)    & 49.2 (0.5)    & 20.3 (0.8)     \\ 
\hline
ENTWSVM                           & Dual                          & quadprog                         & 3.9 (0.0)     & 43.6 (0.6)    & 17.8 (0.4)    & 19.6 (0.2)    & 45.3 (0.6)    & 18.4 (0.3)    & 147.8 (2.4)   & 47.3 (0.4)    & 18.7 (0.4)    & 2627.2 (40.9) & 50.1 (0.3)    & 19.4 (0.3)     \\
ENTWSVM                           & Dual                          & OSQP                             & 1.1 (0.0)     & 43.7 (0.6)    & 18.0 (0.3)    & 2.0 (0.0)     & 44.2 (0.5)    & 18.1 (0.3)    & 4.5 (0.1)     & 46.6 (0.4)    & 20.2 (0.4)    & 16.7 (0.2)    & 49.5 (0.5)    & 19.4 (0.6)     \\
\hline
\end{tabular}}
\caption*{\footnotesize
    \textbf{TABLE NOTE:} The running time is measured in minutes.
    Opt: optimization method. EGKL and TE (misclassification rate) are multiplied by 100; means are reported with standard deviations in parentheses.}
\label{table2:comparison}
\end{table}

\medskip
\noindent
\textbf{Example 2 (Balanced, Dependent Setting)}. 
In this example, the input variables are correlated. Data are generated from a multivariate normal distribution with $\boldsymbol{\mu_{+1}} = [\boldsymbol{1}_{5} \enskip \boldsymbol{0}_{p-5}]^{\small\mathrm{T}}$ and $\boldsymbol{\mu_{-1}} = [-\boldsymbol{1}_{5} \enskip \boldsymbol{0}_{p-5}]^{\small\mathrm{T}}$, and covariance matrix $\bSigma$ as in \eqref{ex2covamax2}, where $\bSigma_{5 \times 5}^*$ has unit diagonal and off-diagonal elements equal to $\rho = 0.8$. The Bayes classifier depends only on the highly correlated variables $x_1, \ldots, x_5$, with a Bayes error of 0.138, independent of $p$.

\begin{equation}\label{ex2covamax2}
\centering
\bSigma =
\begin{bmatrix}
    \bSigma_{5 \cross 5}^*       & \boldsymbol{0}_{5,p-5} \\
    \boldsymbol{0}_{p-5,5}       & \mathbf{I}_{p-5} 
\end{bmatrix}
\end{equation}

\medskip
\noindent
\textbf{Example 3 (Unbalanced, Dependent Setting)}. 
In this example, the relevant variables contribute at different levels to the decision boundary and have unequal pairwise correlations. The classes are imbalanced, with $n_{+1} = \floor{0.6n}$ and $n_{-1} = n - n_{+1}$.
Data are generated from a multivariate normal distribution with $\boldsymbol{\mu_{+1}} = [\boldsymbol{1}_{5} \enskip \boldsymbol{0}_{p-5}]^{\small\mathrm{T}}
$ and $\boldsymbol{\mu_{-1}} = [-\boldsymbol{1}_{5} \enskip \boldsymbol{0}_{p-5}]^{\small\mathrm{T}}
$, and covariance matrix $\bSigma$ as in \eqref{ex2covamax2}, except that $\bSigma_{5 \times 5}^*$ takes the form of \eqref{ex2covamax3} with $\rho = 0.8$. The Bayes classifier depends only on $x_1, \ldots, x_5$, with a Bayes error of 0.115, independent of $p$.

\begin{equation}\label{ex2covamax3}
\centering
\bSigma_{5 \cross 5}^* = 
\begin{bmatrix}
    1 & \rho & \rho^2   & \rho^3 & \rho^4  \\
    \rho & 1 & \rho & \rho^2   & \rho^3  \\
     \rho^2 &  \rho & 1 & \rho & \rho^2 \\
     \rho^3 &  \rho^2 & \rho & 1 & \rho \\
   \rho^4 &  \rho^3 & \rho^2 & \rho & 1 
\end{bmatrix}
\end{equation}

\begin{figure}[H]
	\centering		
		\includegraphics[width=1\textwidth]{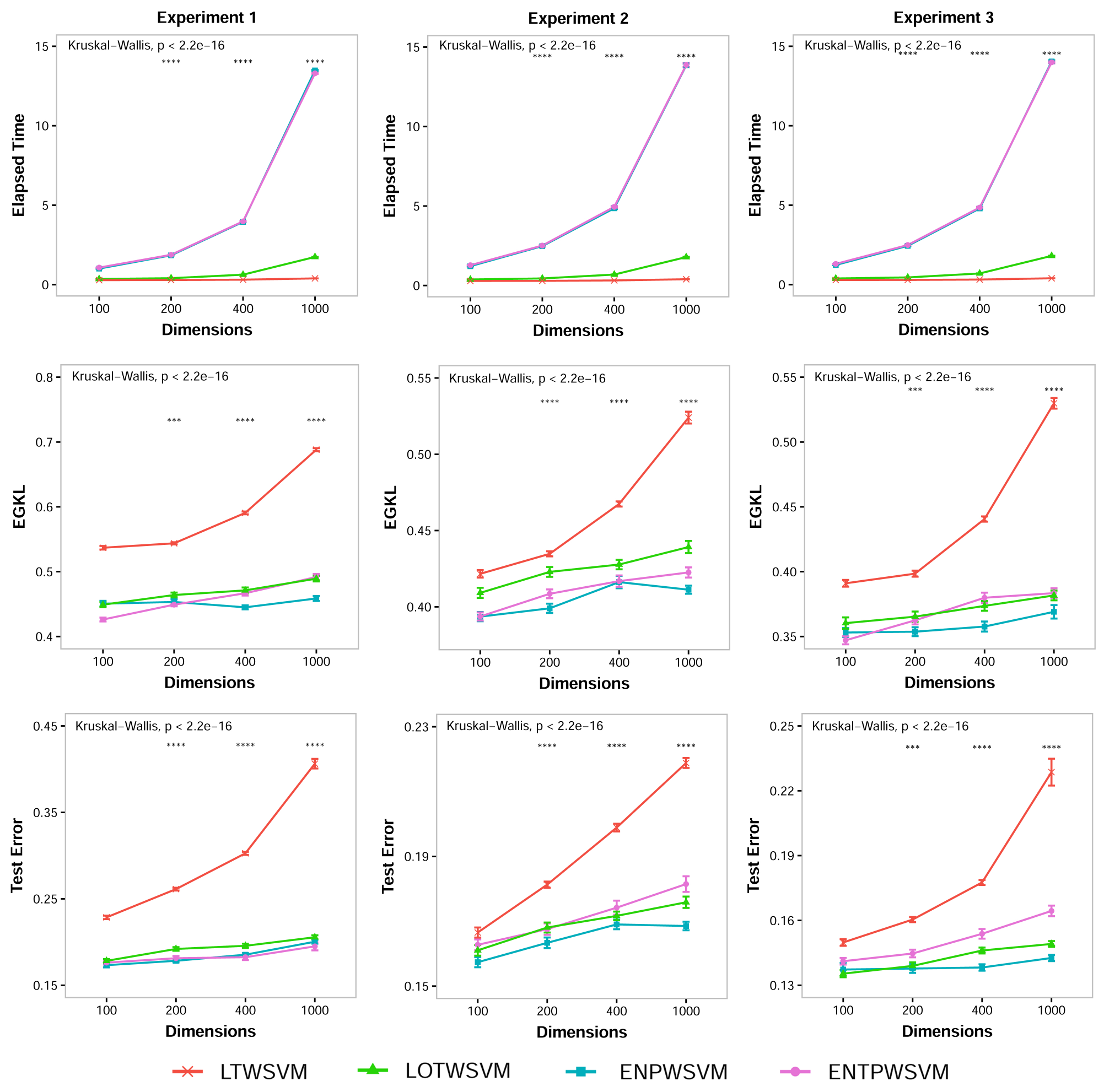}
  \caption{This figure shows performance across Examples 1–3, with colors (or symbols) distinguishing learning methods. Pairwise Wilcoxon signed-rank tests assess significance between dimensions using $p = 100$ as the reference, with stars following R conventions. The global Kruskal–Wallis test evaluates overall performance differences across methods. Data points represent means with standard error bars.}
		\label{fig:anaresult}
\label{fig1:exp_performance}
\end{figure}

Figure \ref{fig1:exp_performance} summarizes the performance of all methods across Examples 1-3 in terms of running time, EGKL (probability estimation accuracy), and test error. Several observations can be made. First, the proposed sparse learning methods LOTWSVM, ENPWSVM, and ENTPWSVM outperform LTWSVM across all examples and dimensions. LTWSVM performs only slightly worse in the low-dimensional case ($n \ge p$) but degrades substantially when $p \gg n$, as it uses all input variables and prediction power is diluted by noise. In contrast, the proposed methods maintain robust performance near the Bayes error as $p$ increases. Second, LOTWSVM achieves comparable accuracy to the elastic net approaches while offering substantially lower computational cost, on a similar scale to LTWSVM in high dimensions. Overall, ENPWSVM as a one-step approach achieves the best performance across all examples.

\begin{figure}[H]
	\centering
		\includegraphics[width=0.95\textwidth]{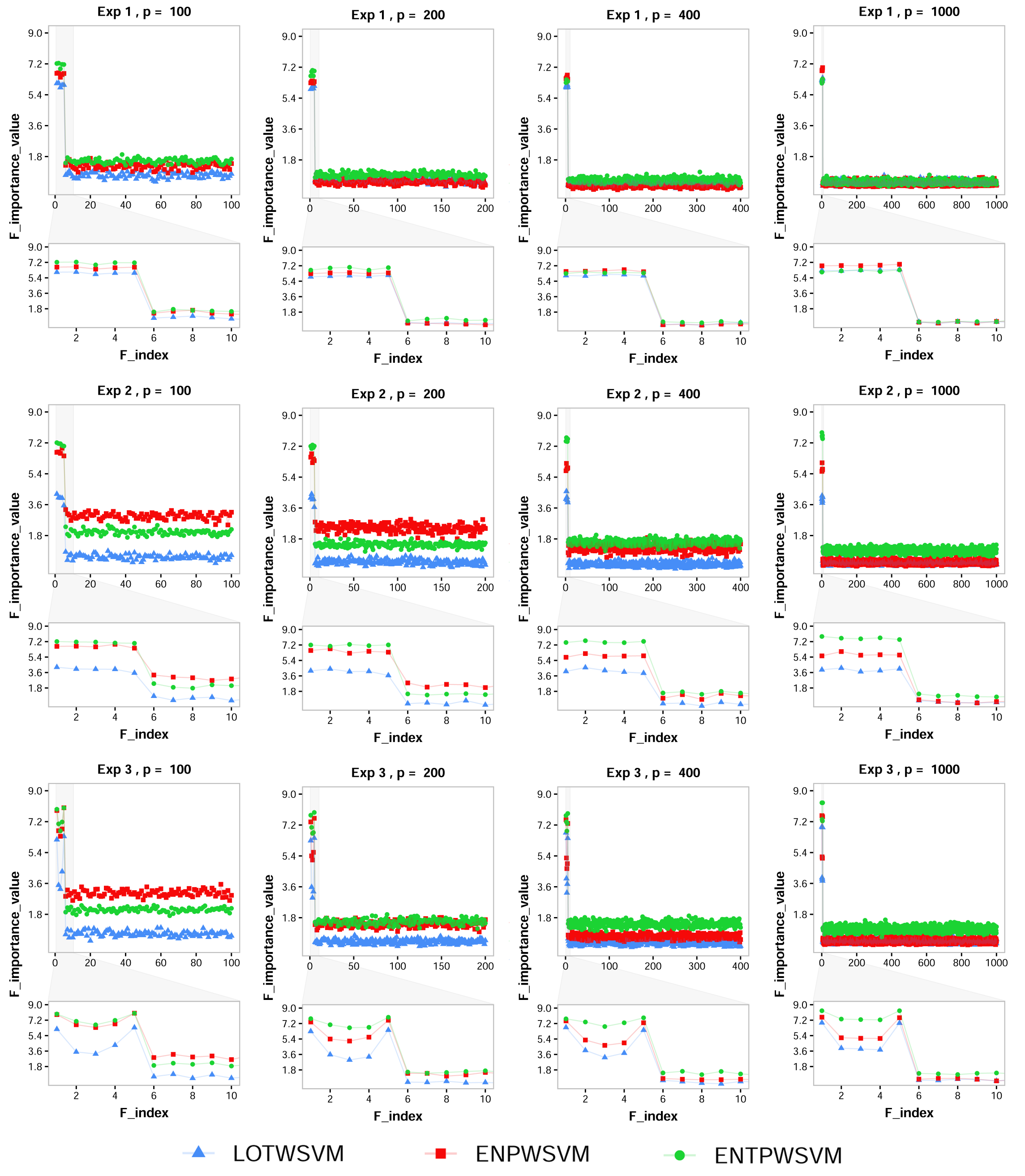}
  \caption{This figure shows feature importance maps for variable selection in Examples 1–3 across various dimensions, with colors (or symbols) distinguishing learning methods. In each subplot, the upper panel displays the full variable frequency map, while the lower panel zooms in on the first 10 variables, including the 5 true features, for clarity.}
		\label{fig:featureimp}
\end{figure}

Beyond class probability estimation and classification, we evaluated the variable selection performance of the proposed sparse learning methods. We define $q_S$ as the number of selected relevant variables (with $p_t=5$
as the oracle in all examples) and $q_N$ as the number of selected noisy variables. The results are summarized in Table \ref{table3:variable_selection}. In Example 1, where variables are independent, both $\ell^1$-norm (LOTWSVM) and elastic net (ENPWSVM/ENTPWSVM) approaches perform similarly, identifying relevant variables and removing most noisy ones. In Examples 2 and 3, where relevant variables are highly correlated, LOTSVM retains only a subset of relevant variables, whereas elastic net-based approaches, particularly the two-step ENTPWSVM, identify all of them due to grouping effects. All proposed methods efficiently remove irrelevant variables. Figure \ref{fig:featureimp} displays feature importance measured by variable selection frequency, confirming that all methods select the correct variables and remove noisy features given an appropriate threshold, outperforming $L_1$-SVM and DrSVM in terms of both classification and variable grouping in \cite{wang_doubly_2006}.

\begin{table}[H]
\caption{Variable selection in Examples 1-3}
\centering
\resizebox{1\textwidth}{!}{
\begin{tabular}{c|ccc|cc|cc|cc} 
\hline

\multirow{2}{*}{\textbf{Methods}} & \multirow{2}{*}{\textbf{n}} & \multirow{2}{*}{\textbf{p}} & \multirow{2}{*}{\textbf{p\textsubscript{t}}} & \multicolumn{2}{c|}{\textbf{Example 1}} & \multicolumn{2}{c|}{\textbf{Example 2}} & \multicolumn{2}{c}{\textbf{Example 3}}  \\ 
\cline{5-10}
                                  &                             &                             &                              & \textbf{q\textsubscript{S}} & \textbf{q\textsubscript{N}}           & \textbf{q\textsubscript{S}} & \textbf{q\textsubscript{N}}          & \textbf{q\textsubscript{S}} & \textbf{q\textsubscript{N}}           \\ 
\hline
LOTWSVM                           & \multirow{3}{*}{100}        & \multirow{3}{*}{100}        & \multirow{3}{*}{5}           & 4.56 (0.10)   & 4.84 (0.77)             & 2.38 (0.09)   & 4.05 (0.85)            & 2.98 (0.09)   & 5.38 (1.10)             \\
~ENPWSVM                          &                             &                             &                              & 4.85 (0.04)   & 11.73 (2.20)            & 4.02 (0.14)   & 32.00 (3.44)           & 4.25 (0.09)   & 33.68 (3.56)            \\
~ENTPWSVM                         &                             &                             &                              & 4.78 (0.06)   & 12.79 (2.17)            & 4.67 (0.07)   & 19.83 (2.42)           & 4.70 (0.06)   & 19.96 (2.55)            \\ 
\hline
LOTWSVM                           & \multirow{3}{*}{100}        & \multirow{3}{*}{200}        & \multirow{3}{*}{5}           & 4.69 (0.10)   & 7.59 (0.71)             & 2.51 (0.10)   & 6.52 (0.99)            & 2.85 (0.10)   & 6.02 (1.09)             \\
~ENPWSVM                          &                             &                             &                              & 4.87 (0.03)   & 7.29 (1.26)             & 3.89 (0.15)   & 53.85 (6.65)           & 3.84 (0.11)   & 29.60 (4.91)            \\
~ENTPWSVM                         &                             &                             &                              & 4.74 (0.07)   & 12.07 (2.44)            & 4.88 (0.04)   & 27.22 (4.21)           & 4.71 (0.07)   & 27.44 (3.81)            \\ 
\hline
LOTWSVM                           & \multirow{3}{*}{100}        & \multirow{3}{*}{400}        & \multirow{3}{*}{5}           & 4.31 (0.15)   & 11.23 (0.95)            & 2.64 (0.11)   & 9.09 (1.01)            & 3.01 (0.09)   & 10.09 (1.22)            \\
~ENPWSVM                          &                             &                             &                              & 4.88 (0.03)   & 9.42 (0.31)             & 3.67 (0.16)   & 55.40 (9.32)           & 3.67 (0.11)   & 28.86 (7.78)            \\
~ENTPWSVM                         &                             &                             &                              & 4.43 (0.10)   & 14.35 (3.39)            & 4.77 (0.06)   & 45.06 (6.18)           & 4.74 (0.07)   & 46.64 (7.19)            \\ 
\hline
LOTWSVM                           & \multirow{3}{*}{100}        & \multirow{3}{*}{1000}       & \multirow{3}{*}{5}           & 4.30 (0.15)   & 19.86 (1.39)            & 2.27 (0.12)   & 15.06 (1.49)           & 3.08 (0.11)   & 17.43 (1.52)            \\
~ENPWSVM                          &                             &                             &                              & 4.91 (0.03)   & 18.97 (0.70)            & 3.66 (0.12)   & 18.02 (2.56)           & 3.82 (0.10)   & 22.47 (4.14)            \\
~ENTPWSVM                         &                             &                             &                              & 4.16 (0.09)   & 16.35 (2.24)            & 4.78 (0.06)   & 56.81 (11.03)          & 4.67 (0.07)   & 51.67 (7.96)            \\
\hline
\end{tabular}}
\label{table3:variable_selection}
\end{table}

\section{Real Data Analysis}\label{sec:realdata}
In this section, we evaluate the proposed methods against two widely used feature selection techniques: Logistic Regression with Adaptive Lasso (LGAL) \citep{li2022logistic} and Cross-Validated XGBoost with Recursive Feature Elimination (RCVXGB) \citep{fuh10433805}, on two real-world datasets: the Leukemia microarray dataset and the MNIST handwritten digits database.

\subsection{Leukemia Microarray Analysis}

The leukemia gene microarray dataset \citep{Golub286}, obtained from \url{https://hastie.su.domains/CASI_files/DATA/leukemia.html}, comprises 3,571 gene expression measurements from 72 leukemia patients: 47 with acute lymphoblastic leukemia (ALL) and 25 with acute myeloid leukemia (AML). We randomly split the data into 38 training samples (25 ALL, 13 AML) and 34 test samples (22 ALL, 12 AML), further dividing training and tuning sets equally over 10 repetitions, and report average test performance with standard error in Table \ref{table4:realexamples}. 

Though less time-efficient, both elastic net-based approaches achieve the best overall performance, with ENTPWSVM excelling in class probability estimation and variable grouping. LOTWSVM runs comparably to LTWSVM while achieving performance close to the elastic net-based methods, selecting fewer genes than the training sample size, a subset of those identified by ENPWSVM and ENTPWSVM. All proposed methods outperform LTWSVM, which lacks variable selection, in both class probability estimation and classification. Similar to LOTWSVM, the $\ell^1$-norm–based LGAL lacks a grouping effect and tends to arbitrarily select one feature from a correlated group while zeroing out the rest. In contrast, RCVXGB with subsampling retains many correlated features, as importance is distributed across similar predictors, exhibiting grouping-like behavior analogous to elastic net-regularized wSVMs. Overall, all wSVM-based methods, including LTWSVM, consistently outperform LGAL and RCVXGB in both class probability estimation and classification accuracy. However, RCVXGB is the most computationally expensive, relying on nested iterative procedures with repeated cross-validation and sequential, non-parallelizable training stages.

\subsection{MNIST Handwritten Digit Recognition}

For the second example, we use the MNIST handwritten digits database (0–9) \citep{LECUN10}, loaded directly from \texttt{Keras} \citep{chollet2015keras}, containing 60,000 training and 10,000 test examples. Each digit is size-normalized and centered as a $28\times28$ grayscale image, yielding 784 features with pixel values scaled to $[0,1]
$. We focus on digit 6 versus digit 9, randomly sampling 250+250 training images from 5,918 sixes and 5,949 nines in the training set, and 750+750 test images from 958 sixes and 1,009 nines in the test set. Training and tuning sets are split equally over 10 repetitions, and average test performance with standard error is reported in Table \ref{table4:realexamples}.

Consistent with the Leukemia results, all proposed methods outperform LTWSVM in both class probability estimation and classification accuracy, with ENTPWSVM achieving the best overall performance and grouping effect. LOTWSVM performs comparably to ENTPWSVM/ENPWSVM at computational cost similar to LTWSVM. LGAL, lacking a grouping mechanism under its $\ell^1$
-norm penalty, exhibits weaker and less stable feature selection among highly correlated pixel features. RCVXGB captures nonlinear structures but yields inferior probability estimation at substantially higher computational cost due to nested cross-validation, mirroring trends from the Leukemia study. LOTWSVM thus provides an effective balance of performance and scalability for high-dimensional settings.

\newcolumntype{Y}{>{\Centering\arraybackslash\hspace{0pt}}X}

\begin{table}[H]
\centering
\caption{Performance Measures for Real Data Sets}
\label{table4:realexamples}
\scriptsize 
\setlength{\tabcolsep}{1.1pt} 
\renewcommand{\arraystretch}{1.0} 
\setlength{\extrarowheight}{-0.5pt}

\begin{tabularx}{\textwidth}{>{\Centering}m{1.5cm}|>{\Centering}p{0.55cm} >{\Centering}p{0.75cm} >{\Centering}p{0.75cm}|>{\Centering}m{0.9cm}|*{4}{Y}|Y|Y}
\hline
\multirow{2}{*}{\textbf{Dataset}} & \multicolumn{3}{c|}{\textbf{Data Info}} & \multirow{2}{*}{\textbf{Eval}} & \multicolumn{6}{c}{\textbf{Methods}} \\ \cline{2-4} \cline{6-11} 
 & \textbf{$n$} & \textbf{$\tilde{n}$} & \textbf{$p$} & & \textbf{LTWSVM} & \textbf{LOTWSVM} & \textbf{ENPWSVM} & \textbf{ENTPWSVM} & \textbf{LGAL} & \textbf{RCVXGB} \\ \hline

\multirow{4}{1.5cm}{\centering Leukemia} & \multirow{4}{*}{38} & \multirow{4}{*}{34} & \multirow{4}{*}{3571} 
& Time & 0.03 (0.00) & 1.00 (0.02) & 45.42 (3.41) & 45.61 (3.62) & 0.01 (0.00) & 73.15 (4.15) \\ 
 & & & & EGKL & 25.99 (2.72) & 7.54 (1.24) & 6.13 (0.41) & 5.66 (0.37) & 55.11 (3.82) & 42.66 (3.12) \\ 
 & & & & \centering TE & 7.35 (1.26) & 0.88 (0.45) & 0.00 (0.00) & 0.00 (0.00) & 11.89 (1.77) & 9.56 (1.53) \\ 
 & & & & $p^*$ & 3571 & 17 & 57 & 84 & 21 & 43 \\ \hline

\multirow{4}{1.5cm}{\centering \makecell{MNIST\\(6 vs 9)}} & \multirow{4}{*}{500} & \multirow{4}{*}{1500} & \multirow{4}{*}{784} 
& Time & 0.33 (0.00) & 0.95 (0.01) & 5.35 (0.04) & 5.87 (0.03) & 0.21 (0.01) & 23.45 (1.32) \\ 
 & & & & EGKL & 10.82 (0.39) & 6.39 (0.11) & 5.56 (0.07) & 5.50 (0.07) & 31.89 (2.76) & 22.66 (2.11) \\ 
 & & & & \centering TE & 1.02 (0.02) & 0.31 (0.01) & 0.10 (0.01) & 0.08 (0.01) & 1.79 (0.03) & 1.42 (0.02) \\ 
 & & & & $p^*$ & 784 & 51 & 139 & 206 & 59 & 117 \\ \hline
\end{tabularx}

\vspace{2pt}
\caption*{\footnotesize
    NOTE: The first column lists the datasets; the next three columns
    report training size $n$, test size $\tilde{n}$, and dimensionality $p$.
    Remaining columns compare the proposed methods and two feature selection approaches: 
    Logistic Regression Adaptive Lasso (LGAL) and XGBoost with RFE and Cross-Validation (RCVXGB). 
    Running time is in minutes; $p^*$ denotes selected features. EGKL and TE (test error) are 
    scaled by 100, with standard errors in parentheses.}
\end{table}

\section{Concluding Remarks}\label{sec:cond}
We proposed $\ell^1$-norm and elastic net regularized wSVM methods for robust probability estimation and variable selection in binary classification. These methods are well-suited for high-dimensional sparse settings, efficiently removing redundant variables and identifying relevant ones. Unlike existing approaches such as $\ell^1$-norm SVM and DrSVM, our methods provide high-confidence class probability estimates with accurate classification and retain the ability to select correlated variable groups. We develop efficient LP and QP algorithms for class probability estimation, compatible with common platforms such as R, Python, and MATLAB, along with tuning parameter selection to obtain optimal results in practice. 

Several directions could extend the sparse learning framework with wSVMs. The proposed methods (LOTWSVM, ENPWSVM, ENTPWSVM) are solved via LP and QP using the \texttt{OSQP} solver in R. Their computational complexity depends on dimensionality $p$, which may become prohibitive for ultra-high-dimensional problems, especially with cross-validation under limited training samples. \cite{Zhu04,wang_doubly_2006} develop efficient solution path algorithms exploiting the piecewise linear structure of $\lambda$
for $\ell^1$-norm and elastic net regularized SVMs. It would be interesting to extend such solution path algorithms to regularized wSVMs for robust class probability estimation in ultra-high dimensions. 

Another promising direction is extending the current binary framework to multiclass probability estimation and variable selection with wSVMs. Many applications, such as image classification and object tracking, including medical image screening with small target regions and substantial noise, e.g., brain tumor MRI image classification \citep{zeng2024102451}, require efficient removal of irrelevant features for accurate classification with reliable confidence measures. \cite{wang_multiclass_2019} and \cite{zeng_wsvms_2022} developed robust multiclass probability estimation algorithms using $L_2$-wSVMs via ensemble learning. Extending the sparse learning framework with wSVMs to multiclass settings is both theoretically interesting and practically important.

\medskip
\begin{center}
{\large\bf SUPPLEMENTARY MATERIAL}
\end{center}
We developed the R package \texttt{BSPwSVMs} for implementing the proposed methods, available on GitHub at \url{https://github.com/zly555/BSPwSVMs}. Usage instructions and a simulation example are provided in the \texttt{README.md} file.


\bibliographystyle{Chicago}
\bibliography{reflist}

@article{lin_support_2002,
	title = {Support Vector Machines for Classification in Nonstandard Situations},
	volume = {46},
	doi = {https://doi.org/10.1023/A:1012406528296},
	language = {en},
	journal = {Machine Learning},
	author = {Lin, Yi and Lee, Yoonkyung and Wahba, Grace},
	year = {2002},
	pages = {191--202},
}

@techreport{dekel_support_2016,
	title = {Support Vector Machines on a Budget},
	language = {en},
	institution = {School of Computer Science and Engineering, The Hebrew University},
	author = {Dekel, Ofer and Singer, Yoram},
	year = {2016},
}

@article{hastie_entire_2004,
	title = {The Entire Regularization Path for the Support Vector Machine},
	volume = {5},
	language = {en},
	journal = {Journal of Machine Learning Research},
	author = {Hastie, Trevor and Rosset, Saharon and Tibshirani, Robert and Zhu, Ji},
	month = oct,
	year = {2004},
	pages = {1391--1415},
}

@article{wang_multiclass_2019,
	title = {Multiclass Probability Estimation With Support Vector Machines},
	volume = {28},
	issn = {1061-8600, 1537-2715},
	url = {https://www.tandfonline.com/doi/full/10.1080/10618600.2019.1585260},
	doi = {10.1080/10618600.2019.1585260},
	language = {en},
	number = {3},
	urldate = {2021-06-03},
	journal = {Journal of Computational and Graphical Statistics},
	author = {Wang, Xin and Zhang, Hao Helen and Wu, Yichao},
	month = jul,
	year = {2019},
	pages = {586--595},
}

@book{scholkopf_learning_2002,
	title = {Learning with Kernels: Support Vector Machines, Regularization, Optimization, and Beyond},
	isbn = {978-0-262-25693-3},
	publisher = {The MIT Press, Cambridge, Massachusetts, USA},
	author = {Schölkopf, Bernhard and Smola, Alexander J.},
	year = {2002},
}

@book{cristianini_shawe-taylor_2000, place={Cambridge}, title={An Introduction to {Support} {Vector} {Machines} and other Kernel-based Learning Methods}, DOI={10.1017/CBO9780511801389}, publisher={Cambridge University Press, Cambridge, UK}, author={Cristianini, Nello and Shawe-Taylor, John}, year={2000}}

@article{Stellato_osqp_20,
  author  = {Stellato, B. and Banjac, G. and Goulart, P. and Bemporad, A. and Boyd, S.},
  title   = {{OSQP}: an operator splitting solver for quadratic programs},
  journal = {Mathematical Programming Computation},
  volume  = {12},
  number  = {4},
  pages   = {637--672},
  year    = {2020},
  doi     = {10.1007/s12532-020-00179-2},
  url     = {https://doi.org/10.1007/s12532-020-00179-2},
}

@article{wang_doubly_2006,
	title = {THE DOUBLY REGULARIZED SUPPORT VECTOR MACHINE},
	volume = {16},
	issn = {10170405, 19968507},
	url = {http://www.jstor.org/stable/24307560},
	number = {2},
	urldate = {2022-07-08},
	journal = {Statistica Sinica},
	author = {Wang, Li and Zhu, Ji and Zou, Hui},
	year = {2006},
	pages = {589--615},
}

@book{Hastie2009,
  author = {Hastie, Trevor and Tibshirani, Robert and Friedman, Jerome},
  edition = 2,
  publisher = {Springer, New York, New York, USA},
  title = {The Elements of Statistical Learning: Data Mining, Inference and Prediction},
  url = {http://www-stat.stanford.edu/~tibs/ElemStatLearn/},
  year = 2009
}

@ARTICLE{KW1971,
  author = {Kimeldorf, G. and Wahba, G.},
  title = {Some results on {T}chebycheffian spline functions.},
  journal = {Journal of Mathematical Analysis and Applications},
  year = {1971},
  volume = {33},
  pages = {82-95}
}

@book{Wahba90,
  author = {Wahba, Grace},
  title = {Spline Models for Observational Data.},
  publisher = {CBMS-NSF Regional Conference Series in Applied Mathematics. SIAM, Philadelphia, Pennsylvania, USA},
  year = {1990}
}

@CONFERENCE{Zhu04,
  author = {Zhu, J. and Rosset, S. and Hastie, T. and Tibshirani, R.},
  title = {1-norm Support Vector Machines},
  booktitle = {The Annual Conference on Neural Information Processing Systems.},
  year = {2004},
  volume = {16}
}

@ARTICLE{WSL2008,
  author = {Wang, J. and Shen, X. and Liu, Y.},
  title = {Probability estimation for large margin classifiers.},
  journal = {Biometrika},
  year = {2008},
  volume = {95},
  pages = {149-167}
}

@article{ye2012,
  title={Sparse methods for biomedical data},
  author={Ye, Jieping and Liu, Jun},
  journal={ACM Sigkdd Explorations Newsletter},
  volume={14},
  number={1},
  pages={4--15},
  year={2012},
  publisher={ACM New York, NY, USA}
}

@article{ye_extension_1989,
	title = {An extension of {Karmarkar}'s projective algorithm for convex quadratic programming},
	volume = {44},
	issn = {0025-5610, 1436-4646},
	url = {http://link.springer.com/10.1007/BF01587086},
	doi = {10.1007/BF01587086},
	language = {en},
	number = {1-3},
	urldate = {2022-03-28},
	journal = {Mathematical Programming},
	author = {Ye, Yinyu and Tse, Edison},
	month = may,
	year = {1989},
	pages = {157--179},
}

@article{zeng_wsvms_2022,
    author = {Liyun Zeng and Hao Helen Zhang},
    title = {Linear Algorithms for Robust and Scalable Nonparametric Multiclass Probability Estimation},
    journal = {Journal of Data Science},
    volume = {21},
    number = {4},
    year = {2022},
    pages = {658--680},
    doi = {10.6339/22-JDS1069},
    issn = {1680-743X},
    publisher = {School of Statistics, Renmin University of China}
}

@book{Bishop1162264,
author = {Bishop, Christopher M.},
title = {Pattern Recognition and Machine Learning (Information Science and Statistics)},
year = {2006},
isbn = {0387310738},
publisher = {Springer, New York, New York, USA}
}

@ARTICLE{Li9088292,
  author={Li, Xiaoping and Wang, Yadi and Ruiz, Rubén},
  journal={IEEE Transactions on Cybernetics}, 
  title={A Survey on Sparse Learning Models for Feature Selection}, 
  year={2022},
  volume={52},
  number={3},
  pages={1642-1660},
  doi={10.1109/TCYB.2020.2982445}}

@article{Pedrobbk007,
    author = {Larrañaga, Pedro and Calvo, Borja and Santana, Roberto and Bielza, Concha and Galdiano, Josu and Inza, Iñaki and Lozano, José A. and Armañanzas, Rubén and Santafé, Guzmán and Pérez, Aritz and Robles, Victor},
    title = "{Machine learning in bioinformatics}",
    journal = {Briefings in Bioinformatics},
    volume = {7},
    number = {1},
    pages = {86-112},
    year = {2006},
    month = {03},
    issn = {1467-5463},
    doi = {10.1093/bib/bbk007},
    url = {https://doi.org/10.1093/bib/bbk007},
    eprint = {https://academic.oup.com/bib/article-pdf/7/1/86/23992771/bbk007.pdf},
}

@INPROCEEDINGS{Wang6126288,
  author={Hua Wang and Nie, Feiping and Huang, Heng and Risacher, Shannon and Ding, Chris and Saykin, Andrew J and Shen, Li},
  booktitle={2011 International Conference on Computer Vision}, 
  title={Sparse multi-task regression and feature selection to identify brain imaging predictors for memory performance}, 
  year={2011},
  volume={},
  number={},
  pages={557-562},
  doi={10.1109/ICCV.2011.6126288}}

@INPROCEEDINGS{Rohini7887916,
  author={Muthukrishnan, R and Rohini, R},
  booktitle={2016 IEEE International Conference on Advances in Computer Applications (ICACA)}, 
  title={LASSO: A feature selection technique in predictive modeling for machine learning}, 
  year={2016},
  volume={},
  number={},
  pages={18-20},
  doi={10.1109/ICACA.2016.7887916}}

@article{Tibshirani02080,
author = {Tibshirani, Robert},
title = {Regression Shrinkage and Selection Via the {Lasso}},
journal = {Journal of the Royal Statistical Society: Series B (Methodological)},
volume = {58},
number = {1},
pages = {267-288},
doi = {https://doi.org/10.1111/j.2517-6161.1996.tb02080.x},
year = {1996}
}

@ARTICLE{Donoho95waveletshrinkage,
    author = {David L. Donoho and Iain M. Johnstone and Gerard Kerkyacharian and Dominique Picard},
    title = {Wavelet shrinkage: asymptopia},
    journal = {Journal of the Royal Statistical Society, Ser. B},
    year = {1995},
    pages = {371--394}
}

@inproceedings{ng_feature_2004,
	address = {Banff, Alberta, Canada},
	title = {Feature selection, \textit{$L_1$} vs. \textit{$L_2$} regularization, and rotational invariance},
	url = {http://portal.acm.org/citation.cfm?doid=1015330.1015435},
	doi = {10.1145/1015330.1015435},
	language = {en},
	urldate = {2022-07-10},
	booktitle = {Twenty-first international conference on {Machine} learning  - {ICML} '04},
	publisher = {ACM Press},
	author = {Ng, Andrew Y.},
	year = {2004},
	pages = {78},
}

@article{reimand_pathway_2019,
	title = {Pathway enrichment analysis and visualization of omics data using g:{Profiler}, {GSEA}, {Cytoscape} and {EnrichmentMap}},
	volume = {14},
	issn = {1750-2799},
	url = {https://doi.org/10.1038/s41596-018-0103-9},
	doi = {10.1038/s41596-018-0103-9},
	number = {2},
	journal = {Nature Protocols},
	author = {Reimand, Jüri and Isserlin, Ruth and Voisin, Veronique and Kucera, Mike and Tannus-Lopes, Christian and Rostamianfar, Asha and Wadi, Lina and Meyer, Mona and Wong, Jeff and Xu, Changjiang and Merico, Daniele and Bader, Gary D.},
	month = feb,
	year = {2019},
	pages = {482--517},
}

@article{Rosset2004BoostingAA,
  title={Boosting as a Regularized Path to a Maximum Margin Classifier},
  author={Saharon Rosset and Ji Zhu and Trevor J. Hastie},
  journal={Journal of Machine Learning Research},
  year={2004},
  volume={5},
  pages={941-973}
}

@book{boyd2004convex,
  title={Convex Optimization},
  author={Boyd, Stephen and Vandenberghe, Lieven},
  year={2004},
  publisher={Cambridge University Press, Cambridge, UK}
}

@article{WANG2022382,
title = {A safe double screening strategy for elastic net support vector machine},
journal = {Information Sciences},
volume = {582},
pages = {382-397},
year = {2022},
issn = {0020-0255},
doi = {https://doi.org/10.1016/j.ins.2021.09.026},
url = {https://www.sciencedirect.com/science/article/pii/S0020025521009579},
author = {Hongmei Wang and Yitian Xu}
}

@INPROCEEDINGS{Eladio5400391,
  author={Díaz, Eladio Rodríguez and Castañón, David A.},
  booktitle={Proceedings of the 48h IEEE Conference on Decision and Control (CDC) held jointly with 2009 28th Chinese Control Conference}, 
  title={Support vector machine classifiers for sequential decision problems}, 
  year={2009},
  volume={},
  number={},
  pages={2558-2563},
  doi={10.1109/CDC.2009.5400391}}

@inproceedings{TanWT10,
  author={Mingkui Tan and Li Wang and Ivor W. Tsang},
  title={Learning Sparse SVM for Feature Selection on Very High Dimensional Datasets},
  year={2010},
  cdate={1262304000000},
  pages={1047-1054},
  url={https://icml.cc/Conferences/2010/papers/227.pdf},
  booktitle={ICML},
}

@article{ghaddar_high_2018,
	title = {High dimensional data classification and feature selection using support vector machines},
	volume = {265},
	issn = {0377-2217},
	url = {https://www.sciencedirect.com/science/article/pii/S0377221717307713},
	doi = {https://doi.org/10.1016/j.ejor.2017.08.040},
	number = {3},
	journal = {European Journal of Operational Research},
	author = {Ghaddar, Bissan and Naoum-Sawaya, Joe},
	year = {2018},
	pages = {993--1004},
}

@article{ye_complexity_1998,
	title = {On the complexity of approximating a {KKT} point of quadratic programming},
	volume = {80},
	issn = {1436-4646},
	url = {https://doi.org/10.1007/BF01581726},
	doi = {10.1007/BF01581726},
	number = {2},
	journal = {Mathematical Programming},
	author = {Ye, Yinyu},
	month = jan,
	year = {1998},
	pages = {195--211},
}

@inproceedings{Tseng1988ASP,
  title={A simple polynomial-time algorithm for convex quadratic programming},
  author={Paul Tseng},
  year={1988},
  publisher={Laboratory for Information and Decision Systems, MIT, Cambridge, Massachusetts, USA}
}

@INPROCEEDINGS{Vaidya63499,
  author={Vaidya, P.M.},
  booktitle={30th Annual Symposium on Foundations of Computer Science}, 
  title={Speeding-up linear programming using fast matrix multiplication}, 
  year={1989},
  volume={},
  number={},
  pages={332-337},
  doi={10.1109/SFCS.1989.63499}}

@article{Boyd220016,
author = {Boyd, Stephen and Parikh, Neal and Chu, Eric and Peleato, Borja and Eckstein, Jonathan},
title = {Distributed Optimization and Statistical Learning via the Alternating Direction Method of Multipliers},
year = {2011},
issue_date = {January 2011},
publisher = {Now Publishers Inc.},
address = {Hanover, MA, USA},
volume = {3},
number = {1},
issn = {1935-8237},
url = {https://doi.org/10.1561/2200000016},
doi = {10.1561/2200000016},
journal = {Found. Trends Mach. Learn.},
month = {jan},
pages = {1–122},
numpages = {122}
}

@misc{Berkelaar2004,
  author = {Berkelaar, Michel and Eikland, Kjell and Notebaert, Peter},
  month = {May 1},
  title = {{lp\_solve} 5.5, Open source (Mixed-Integer) Linear Programming system},
  url = {http://lpsolve.sourceforge.net/5.5/},
  year = 2004
}

@article{
Golub286,
author = {T. R. Golub  and D. K. Slonim  and P. Tamayo  and C. Huard  and M. Gaasenbeek  and J. P. Mesirov  and H. Coller  and M. L. Loh  and J. R. Downing  and M. A. Caligiuri  and C. D. Bloomfield  and E. S. Lander },
title = {Molecular Classification of Cancer: Class Discovery and Class Prediction by Gene Expression Monitoring},
journal = {Science},
volume = {286},
number = {5439},
pages = {531-537},
year = {1999},
doi = {10.1126/science.286.5439.531},
URL = {https://www.science.org/doi/abs/10.1126/science.286.5439.531}}

@article{zhang_gene_2006,
	title = {Gene selection using support vector machines with non-convex penalty.},
	volume = {22},
	issn = {1367-4803},
	doi = {10.1093/bioinformatics/bti736},
	language = {eng},
	number = {1},
	journal = {Bioinformatics},
	author = {Zhang, Hao Helen and Ahn, Jeongyoun and Lin, Xiaodong and Park, Cheolwoo},
	month = jan,
	year = {2006},
	pmid = {16249260},
	pages = {88--95},
}

@misc{LECUN10,
author={LeCun, Yann and Cortes, Corinna and  Burges, Christopher J.C.},
  title = {THE {MNIST} DATABASE of handwritten digits},
  howpublished = {\url{http://yann.lecun.com/exdb/mnist/}},
  year = 1998
}

@misc{chollet2015keras,
  title={Keras},
  author={Chollet, Fran\c{c}ois and others},
  year={2015},
  howpublished={\url{https://keras.io}},
}

@article{li2022logistic,
  title={Logistic regression with adaptive sparse group lasso penalty and its application in acute leukemia diagnosis},
  author={Li, Juntao and Liang, Ke and Song, Xuekun},
  journal={Computers in Biology and Medicine},
  volume={141},
  pages={105154},
  year={2022},
  publisher={Elsevier}
}

@article{zeng2024102451,
title = {Robust brain {MRI} image classification with {SIBOW-SVM}},
journal = {Computerized Medical Imaging and Graphics},
volume = {118},
pages = {102451},
year = {2024},
issn = {0895-6111},
doi = {https://doi.org/10.1016/j.compmedimag.2024.102451},
url = {https://www.sciencedirect.com/science/article/pii/S0895611124001289},
author = {Liyun Zeng and Hao Helen Zhang}
}

@INPROCEEDINGS{fuh10433805,
  author={Fuhnwi, Gerard Shu and Revelle, Matthew and Izurieta, Clemente},
  booktitle={2024 IEEE 3rd International Conference on AI in Cybersecurity (ICAIC)}, 
  title={Improving Network Intrusion Detection Performance : An Empirical Evaluation Using Extreme Gradient Boosting (XGBoost) with Recursive Feature Elimination}, 
  year={2024},
  volume={},
  number={},
  pages={1-8},
  doi={10.1109/ICAIC60265.2024.10433805}}

\end{document}